\documentclass{article}
\usepackage[utf8]{inputenc}
\usepackage[english]{babel}

\usepackage{graphicx,soul}
\usepackage{tcolorbox}
\usepackage{wrapfig}
\usepackage{amssymb}
\usepackage{gensymb}
\usepackage{booktabs}
\usepackage{amsmath}
\usepackage{bm}
\usepackage{geometry}
\usepackage{caption}
\usepackage{subcaption}
\usepackage{multirow}
\usepackage{tabularx}
\usepackage[sort,numbers, compress]{natbib}
\usepackage{doi}
\usepackage{hyperref}
\usepackage{cancel}
\usepackage{overpic}

\usepackage[export]{adjustbox}

\usepackage{authblk}

\usepackage{etoolbox}
\makeatletter
\providecommand{\subtitle}[1]{
  \apptocmd{\@title}{\par {\large #1 \par}}{}{}
}
\makeatother

\usepackage{hyperref}
\hypersetup{
    colorlinks=true,
    linkcolor=blue,
    filecolor=blue,      
    urlcolor=blue,
    citecolor=blue
}

\usepackage{longtable}

\usepackage{xcolor}

\usepackage{float}

\usepackage{parskip}

\usepackage{lipsum}

\makeatletter
\renewcommand{\maketitle}{\bgroup\setlength{\parindent}{0pt}
\begin{flushleft}
  {\LARGE \textbf{\@title}} 

  {\small \@author}
\end{flushleft}\egroup
}
\makeatother

\title{Multiscale patterns of migration flows in Austria: regionalization,
  administrative barriers, and urban-rural divides}

\author[1,2,*]{Thomas Robiglio}
\author[2,*]{Martina Contisciani}
\author[2,3]{Márton Karsai}
\author[1,2]{Tiago P. Peixoto}

\affil[1]{Inverse Complexity Lab, IT:U Interdisciplinary Transformation University Austria, 4040 Linz, Austria}
\affil[2]{Department of Network and Data Science, Central European University, 1100 Vienna, Austria}
\affil[3]{National Laboratory for Health Security, HUN-REN Alfréd Rényi Institute of Mathematics, Budapest, 1053, Hungary}

\affil[*]{These authors contributed equally to this work}

\begin{document}
\maketitle

\begin{abstract}
Migration is central in various societal problems related to socioeconomic development. While much of the existing research has focused on international migration, migration patterns within a single country remain relatively unexplored. In this work we study internal migration patterns in Austria for a period of over 20 years, obtained from open and high-granularity administrative records. We employ inferential network methods to characterize the flows between municipalities and extract their clustering according to similar target and destination rates. Our methodology reveals significant deviations from commonly assumed relocation patterns modeled by the gravity law. At the same time, we observe unexpected biases of internal migrations that leads to less frequent movements across boundaries at both district and state levels than predictions suggest. This leads to significant regionalization of migration at multiple geographical scales and augmented division between urban and rural areas. These patterns appear to be remarkably persistent across decades of migration data, demonstrating systematic limitations of conventionally used gravity models in migration studies. Our approach presents a robust methodology that can be used to improve such evaluations, and can reveal new phenomena in migration networks.
\end{abstract}

\section{Introduction}
Migration plays a central role in urbanization, segregation, gentrification, and
in many phenomena related to socioeconomic
development~\cite{papademetriou1991unsettled}. The driving forces of migration
can be extremely varied, including labor market imbalances, wealth inequalities,
conflicts, ethnoracial segregation---reflecting the rapid increase in complexity
of human societies~\cite{czaika2022migration}. Migration has increasingly become
a crucial topic in public discourse and in regional and national governance,
drawing considerable attention in academic research. In recent years,
researchers from various fields have investigated why people migrate, how
migration takes place, and what the consequences of migration are in a broad
sense, both for the migrants themselves and for the communities
involved~\cite{scholten2022introduction}. However, much of this research has
mainly focused on international migration, leaving many questions unanswered for
a comprehensive understanding of internal (or domestic) migration. Since the
majority of migration events in the world occur within national borders, an
emphasis on international migrations overlooks a significant portion of the
overall migration phenomenon. Additionally, focusing on international migrations
restricts our ability to fully assess social dynamics and the impacts of policy
interventions at the local level. To address these gaps, we analyze internal
migration data from Austria spanning a 20-year period, providing a mesoscopic
analysis of domestic relocation patterns.

A common approach for analyzing migration data---and human mobility in
general---is to develop statistical models that allow researchers to uncover the
underlying forces driving these phenomena, make predictions, and test
hypotheses~\cite{barbosa2018human}. At an individual level, these include random
walk frameworks to characterize mobility
trajectories~\cite{brockmann2006scaling, gonzalez2008understanding,
  song2010modelling}. At a population level, the usual task is to model
collective flows between locations, based on attributes such as population
densities, geographic distance, and socio-economic indicators. The most commonly
used modeling ansatz for this purpose is based on the pervasive empirical
observation that the rate of movement between two regions tends to increase
according to the product of their population densities, and inversely
proportional to a function of their distance~\cite{zipf1946the}, which leads to
a family of so-called ``gravity'' models~\cite{lewer2008gravity,
  prieto2018gravity, cabanas2025human}, due the functional similarity with
Newton's law of gravitation. This pattern is observed in many different kinds of
mobility and human behavior, including commutes~\cite{mazzoli2019field,
  kwon2023multiple}, daily movements~\cite{liu2014uncovering, li2021gravity},
mobile phone communications~\cite{expert2011uncovering}, and
migration~\cite{zipf1946the, poot2016gravity}. The simple form of this model
family allows for an intuitive interpretation in terms of the overall
attractivity of each region, which is reasonably correlated to its population,
and the cost of movement, which should grow with distance. However, the precise
functional shape of gravity models, and in particular their parameters, are not
derived from first principles, or from more fundamental processes of mobility,
rendering them effective, rather than mechanistic in nature. Important
alternatives to gravity models do exist, which attempt to include more
mechanistic elements, such as the ``intervening opportunities''
model~\cite{stouffer1940intervening,akwawua2000intervening}, which discards the
explicit dependence on distance, in favor of the cumulative number of
``opportunities'' between source and destination, which although it typically
increases with distance, it can also be associated with other factors. Another,
more recent alternative is the radiation model~\cite{simini2012universal}, which
posits, similarly, that a traveler ranks the available opportunities according
to their distance, and chooses the closest one above a certain fitness
threshold. In general, however, most available empirical mobility data do not
offer more statistical evidence for these alternative models when compared to
the gravity ansatz, which often has a better quality of fit and is more
predictive~\cite{lenormand2016systematic, mazzoli2019field}, although not by
large margins.

Due to the effective and heuristic nature of gravity models, they are not
expected to saturate the modeling requirements of mobility data, and in fact
many systematic deviations have been observed in previous studies, such as
the inability to model behaviors for long distances~\cite{lenormand2016systematic,
  yang2014limits, simini2012universal}, and the overdispersion commonly seen in
the data~\cite{burger2009specification, masucci2013gravity, beyer2022gravity},
when compared to model predictions.

In this work, we are interested in further modeling migration dynamics and
evaluating the gravity model, but from the point of view of arbitrary geographic
discrepancies that are not \emph{a priori} posited explicitly, but instead are
discovered \emph{a posteriori} from data. We do so by leveraging an inferential
network clustering method~\cite{peixoto_bayesian_2019,peixoto2018nonparametric},
that fits a nonparametric generative model to empirical data, instead of
enforcing a particular functional shape for the migration rates. Our model
approximates any functional shape by separating the regions into discrete
groups, and therefore their corresponding migration rates as piecewise mixtures
of elementary distributions. Our procedure includes Bayesian regularization
based on the minimum description length principle~\cite{grunwald2007minimum,
  rissanen2010information}, and hence is robust against overfitting. In
addition, our model works in a hierarchical
fashion~\cite{peixoto_hierarchical_2014}, so that the mixing patterns between
geographical regions are simultaneously modeled at multiple scales.

Based on this inferential analysis, we compare the properties of the
inferred nonparametric and multiscale model to a fitted gravity model, which
then allows us to identify statistically significant discrepancies.

When applying our method to internal migration data for Austria, interestingly, we find systematic biases in migration flows where administrative boundaries appear as effective barriers for relocation. These biases lead to clusters in the migration network that resemble administrative borders between federal states and districts with a remarkable accuracy. This indicates a reduced importance of distance in determining internal migration events, in addition to strong regionalization and elevated urban-rural divide as compared to gravity model predictions.

\section{Results}

Our analysis examines the network of internal migration in Austria in the period
from 2002 to 2021, where each node represents a municipality, and each directed
edge denotes the number of individuals relocating from one municipality to
another. We aggregate the data annually and analyze twenty distinct networks
that capture migration flows for each year. Alternatively, one could consider
combining all data into a single aggregated network. However, such an approach
would hinder our ability to evaluate the consistency of migration patterns and
the robustness of our findings over time. Additional information on the data,
such as the number of nodes and edges for each year, is provided in
Sec.~\ref{sec:data}

Similarly to what is observed in a variety of other mobility data, migration
flows between two locations tend to decay with the distance between them, and
increase with the population densities of the source and target of the migration---presumably due to the increased cost of farther relocations, and the
opportunities associated with moving to higher population areas. This is
typically modeled via a gravity model, where the expected number of
migrations $x_{ij}$ from locations $j$ to $i$ is given by
\begin{equation}
\label{eq:gravity}
   \left<x_{ij}\right>=\mu_{ij}=K\frac{\left(p_ip_j\right)^{\alpha}}{d_{ij}^{\beta}},
\end{equation}
where $p_{i}$ is the population of location $i$, and $d_{ij}$ is the distance
between locations $i$ and $j$. The remaining parameters $K$, $\alpha$, and
$\beta$ are to be inferred from data, since they are context-dependent.
We use a Bayesian approach to infer the unknown parameters by sampling from their joint posterior distribution using Hamiltonian Monte Carlo~\cite{betancourt2013hamiltonianmontecarlohierarchical, gelman2013bayesian}. For further details, we refer to Sec.~\ref{sec:methods}

\begin{figure}
    \centering
    \includegraphics[width=0.4\linewidth]{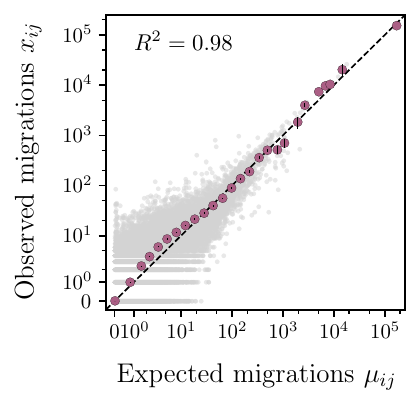}
    \caption{Observed migration counts $x_{ij}$ versus expected counts
      $\mu_{ij}$ according to the inferred gravity model, between municipalities
      in Austria in 2013. The red circles show average values for equal sized
      bins, along with their standard error. The dashed line represents the
      $x_{ij}=\mu_{ij}$ diagonal slope. The coefficient of determination $R^2$ is shown for the fitted model.}
    \label{fig:gravity}
\end{figure}

\begin{figure}[ht!]
    \centering
    \includegraphics[width=0.7\linewidth]{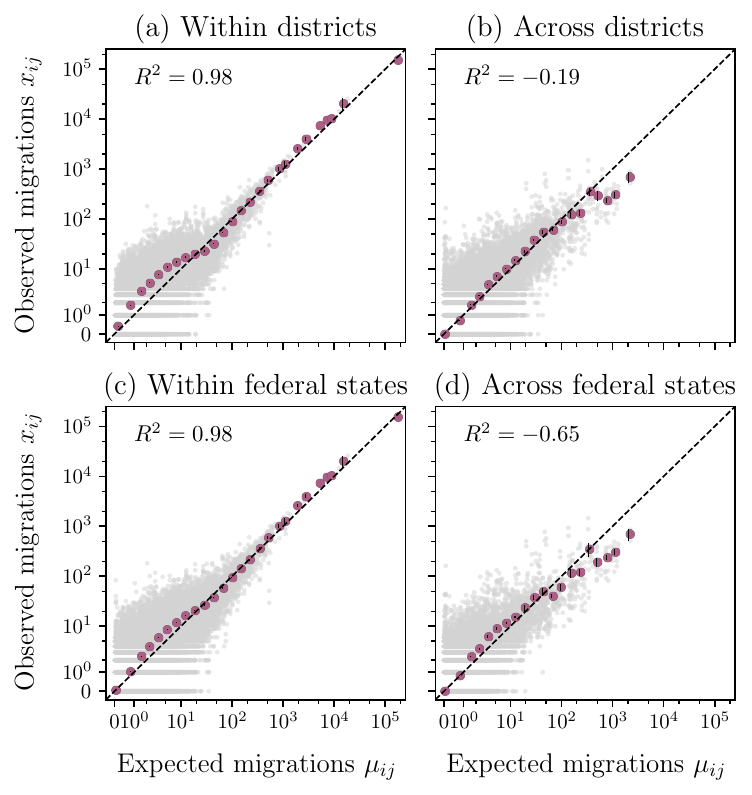}
    \caption{Same analysis of Fig.~\ref{fig:gravity}, but containing only
      migrations within a single administrative region (left panels) and across
      two or more (right panels). The top panels consider districts, and the
      bottom panels federal states.  The expected migrations are inferred from the entire data (i.e., the same as in Fig.~\ref{fig:gravity}), and therefore are the same for all panels, instead of being fitted individually. Likewise, the coefficient of determination $R^2$ for each panel is calculated with respect to the same overall model fitted on the whole data. Hence, a negative value is possible, and indicates a situation where the mean of the data provides a better fit than the function considered.}
    \label{fig:gravity-border}
\end{figure}

In Fig.~\ref{fig:gravity} we show a comparison between the observed and expected
migration counts between municipalities in Austria in 2013, according to the
inferred gravity model. From this type of analysis, which is typically performed
in mobility studies, one could conclude that the gravity model offers a fairly
good fit to the data, with the only noticeable deviation being a systematic
discrepancy for low values of $\mu_{ij}$ in the range between $3$ to $20$
(corresponding to movements between large distances and/or low populations
areas), where more migration events are observed than would be expected by the
gravity model. Already anticipating one of our results obtained via
nonparametric inference, further insight into this discrepancy can be obtained
by stratifying the migrations in two groups, according to whether or not an
administrative boundary has been crossed. In Fig.~\ref{fig:gravity-border} we
show this stratification according to the boundaries between district and
federal states. In both cases we see that the discrepancies for low $\mu_{ij}$
values are exacerbated within administrative boundaries (Fig.~\ref{fig:gravity-border}a and c), and vanish when these
are crossed (Fig.~\ref{fig:gravity-border}b and d). Instead, in the latter case, deviations for high $\mu_{ij}$ are
seen (i.e.\ between short distances and/or high population areas), where the
observed migrations are fewer than expected according to the gravity ansatz.

If the gravity model would hold uniformly for all migrations between
municipalities, this discrepancy would not be expected. However, since the model
is based on a very specific relationship between distance and population, it is
difficult to interpret these deviations, since not only these two quantities
cannot be disentangled, but also the model is incapable of incorporating other
types of geographic dependencies.

To better understand the deviations from the gravity model, we relax the
assumptions of Eq.~\eqref{eq:gravity}, and consider instead a more general model
based on a weighted stochastic block model (WSBM)
formulation~\cite{peixoto2018nonparametric}. More specifically, given a
partition $\bm b$ of the municipalities into $B$ groups, where
$b_{i}\in \{0,\dots,B-1\}$ is the group membership of municipality $i$, we model
the number of migrations between municipalities according to
\begin{equation}\label{eq:model}
  P(\bm x | \bm \theta, \bm b) = \prod_{ij}P(x_{ij}|\theta_{b_{i}, b_{j}}),
\end{equation}
where $P(x|\theta_{r,s})$ is a kernel distribution (e.g. Poisson or geometric)
with parameters $\theta_{rs}$ that are conditioned only the groups $r$ and $s$
of the edge endpoints. Therefore, by choosing this kernel distribution, the
number of groups, and how the nodes are partitioned into it, we can approximate
a wide variety of distributions, including those that deviate significantly from
the gravity model. Importantly, we do not require the municipalities to be
grouped in spatially contiguous regions, and do not enforce any dependency of
the migration rates with distance or population. If such dependencies actually
do exist in the data, we should be able to detect them, but we allow for any
mixing pattern between groups of municipalities to be uncovered as well. This
can be contrasted with most descriptive community detection
methods~\cite{peixoto_descriptive_2023}, which not only expect and enforce a
larger number of connections between nodes of the same group than otherwise
(independently of whether this is the most salient or even statistically valid pattern
in the data~\cite{zhang_statistical_2020}), but they are also unaffected by
mixing preferences between nodes of different groups.

In order to perform the inference of the model of Eq.~\eqref{eq:model}, we employ
the Bayesian approach described in Ref.~\cite{peixoto2018nonparametric}, which
determines the most appropriate model according to the minimum description
length (MDL) principle~\cite{grunwald2007minimum, rissanen2010information}, that
provides robustness against overfitting. One important aspect of our method is
that it yields a hierarchical partition, i.e. the groups themselves are
clustered into their own groups, and so on recursively, yielding a description
of the data in multiple scales. For further details on the methodology, we refer
to Sec.~\ref{sec:methods}

\begin{figure}[]
  \centering
  Migration flows\\\vspace{1em}

  \includegraphics[width=\linewidth]{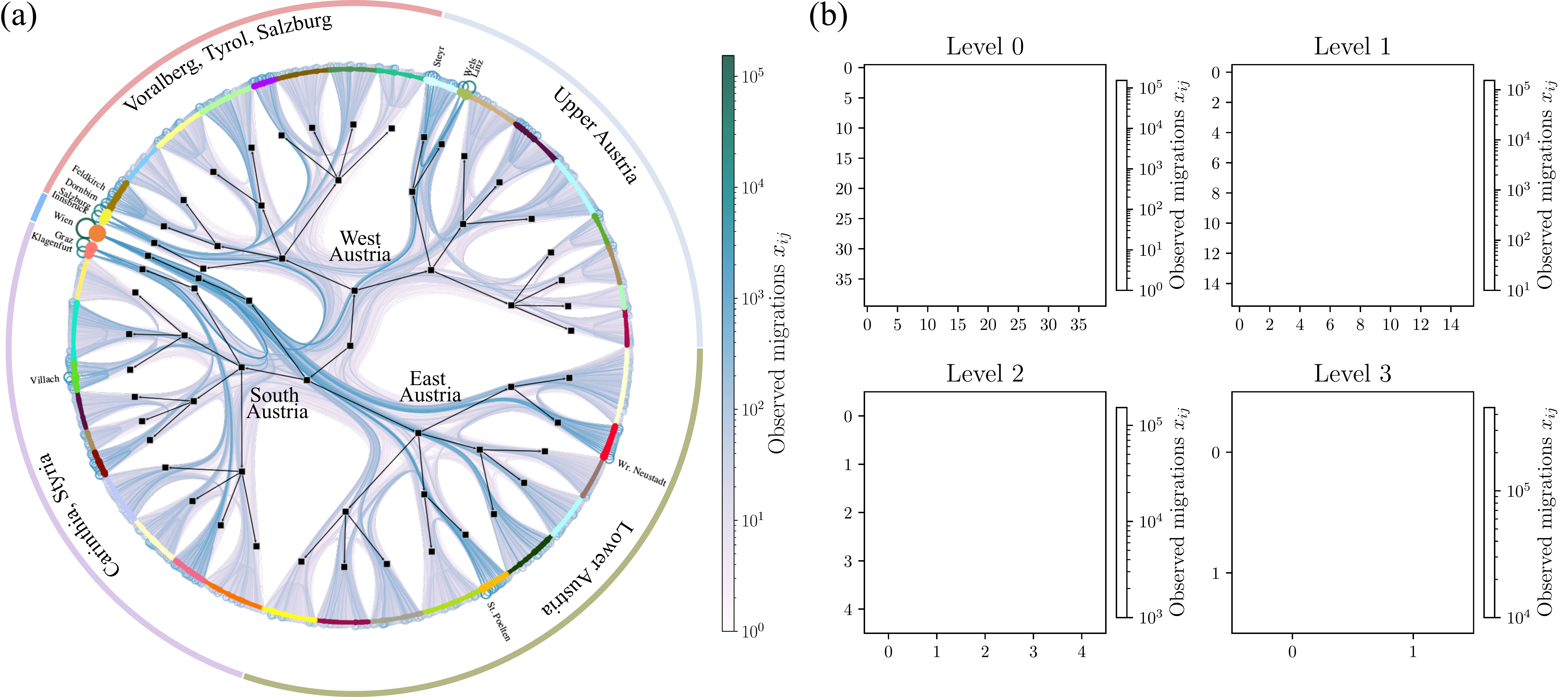}

  Deviations from the gravity model\\\vspace{1em}
  \includegraphics[width=\linewidth]{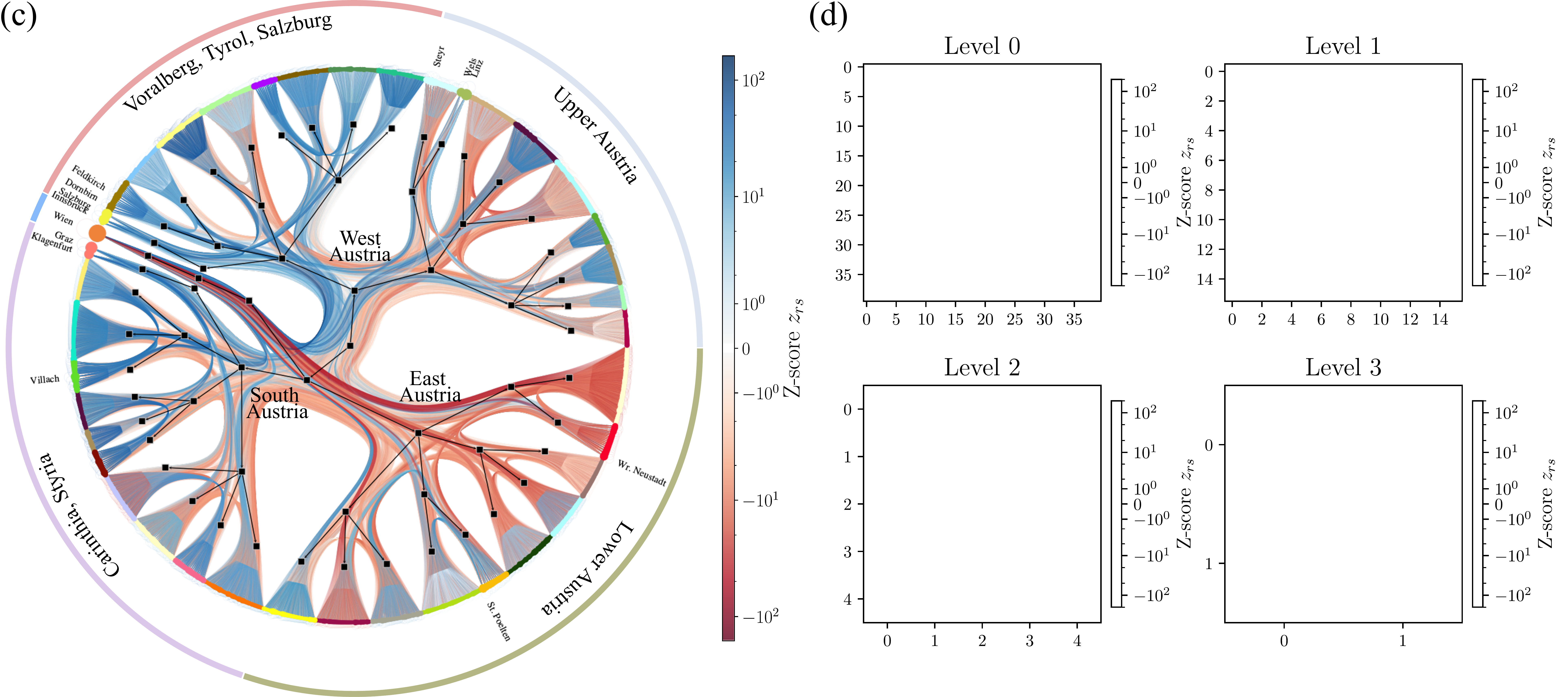}

  \caption{(a) Fit of the WSBM for Austrian migrations in 2013. The edges are
    routed according to the inferred hierarchy (shown in black), using an
    edge-bundling layout~\cite{holten2006hierarchical}. The color of the edges
    indicates their count of migration events in the year. The size of the nodes
    is proportional to their population. The color of the nodes corresponds to
    the partition they belong to at hierarchical level $l=0$. The partitions
    at level $l=2$ are shown with the outer circular arcs. Cities names are
    shown for the 13 largest cities. Levels 2 and 3 are annotated with the most
    frequent federal state classifications (NUTS2), and general areas (NUTS1),
    respectively. (b) Corresponding matrices of migration counts between groups
    for the four different levels of the hierarchy.
    The bottom panels (c) and (d) correspond to (a) and (b), but they show instead the group-level deviation from the gravity model via the z-score, as defined in Eq.~\eqref{eq:zscore}. 
    Blue edges indicate positive discrepancies, meaning that more migration events are observed than predicted by the gravity model. Conversely, red edges represent negative deviations, where the observed migrations are fewer than expected.}
    \label{fig:fig1}
\end{figure}

In Fig.~\ref{fig:fig1}a we show the hierarchical partition of the internal
migration network in 2013, which reveals a heterogeneous group structure,
largely aligned with wider geographical regions at the upper levels. We
emphasize that our model does not \emph{a priori} enforce or expect this
pattern, and the fact it is found means migration is strongly regionalized, both
at a local and global scales. The migration volumes between municipalities vary
by several orders of magnitude, with intra-municipality relocations (i.e.,
self-loops in Fig.~\ref{fig:fig1}a, or the diagonal entries in \ref{fig:fig1}b)
being the most numerous, with a migration rate that increases with the
respective populations. This observation is compatible with the gravity model
ansatz, since these are the relocations with the shortest distances, and also
between the largest population densities in the case of some municipalities.

\begin{figure}
    \centering
    \includegraphics[width=\linewidth]{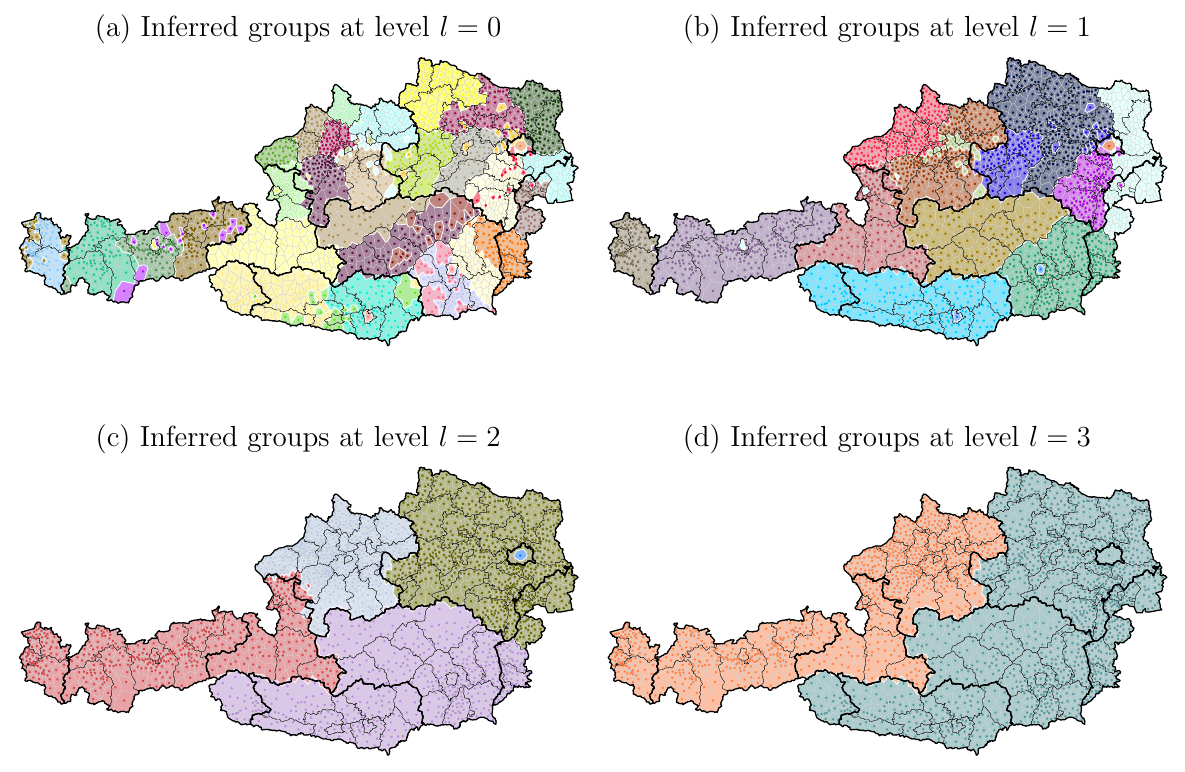}
    \caption{Inferred groups with the WSBM for 2013, for different hierarchical
      levels $l$. Thick black lines indicate federal states boundaries, thin
      black lines denote districts borders. }
    \label{fig:partition}
\end{figure}

However, what is well captured by our WSBM, but not so much by the gravity model, is
the strength and specificity of the regionalization, with the overall regions and
federal states being very well captured by the hierarchical partition.
This becomes especially evident when the inferred partitions are overlaid on the map of Austria, as shown in Fig.~\ref{fig:partition}.
Furthermore, the lowest inferred hierarchical level (Fig.~\ref{fig:partition}a) also separates the larger
cities and urban areas from the rest, which act as source and destination hubs
for migrations to and from both large and small municipalities, as well as
nearby and distant locations. This is more clearly illustrated in
Fig.~\ref{fig:fig1}b, where the communities containing Wien, Graz, and Salzburg
(respectively, communities 0, 1, and 39, at level $l=0$) show large out-diagonal
entries, indicating frequent interactions with all other communities.
Additionally, this plot highlights that the overall mixing pattern is highly
assortative at all levels, meaning that individuals are more likely to migrate
within the same cluster, and the same sub-cluster inside that cluster, which
often corresponds to a specific geographical region.

In the bottom panels of Fig.~\ref{fig:fig1} we show also the deviation of the
observed flows between groups from what would be expected with the gravity
model, via the z-score between groups $r$ and $s$ defined as
\begin{equation}\label{eq:zscore}
  z_{rs} = \frac{x_{rs} - \mu_{rs}}{\sqrt{\mu_{rs}}},
\end{equation}
where $x_{rs} = \sum_{ij}x_{ij}\delta_{b_i,r}\delta_{b_j,s}$ and
$\mu_{rs} = \sum_{ij}\mu_{ij}\delta_{b_i,r}\delta_{b_j,s}$, and we used the fact
that the sum of independent Poisson variables is also Poisson-distributed with
the summed means. A positive or negative value with magnitude $|z_{rs}|>3$ is
considered significant, and means that the gravity model formulation cannot
plausibly account for the deviation. 
Specifically, positive z-scores (shown in blue) imply that the observed migrations exceed the model' expectations---i.e., the gravity model underestimates the flows. Conversely, negative values (shown in red) indicate an overestimation by the model.
As we see in Fig.~\ref{fig:fig1}c, relative
to the gravity model, there are fewer migration events between many nearby
regions of Lower Austria and Vienna, which is compensated for a relative
abundance of migrations from farther regions in Voralberg, Tyrol, and Salzburg.
Most internal migrations between inferred clusters tend to exceed the
expectation from the gravity model, with a few exceptions where the opposite is
observed.

\begin{figure}[h!]
  \centering
    Full migration network\\\vspace{1em}
   \includegraphics[width=\linewidth]{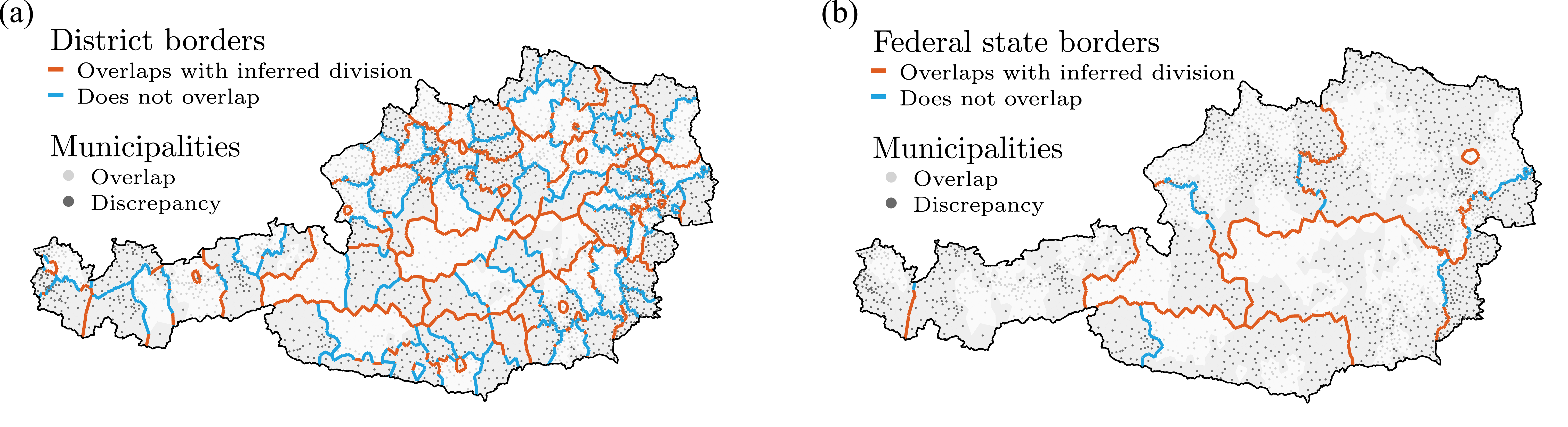}
  
      Binarized network\\\vspace{1em}
   \includegraphics[width=\linewidth]{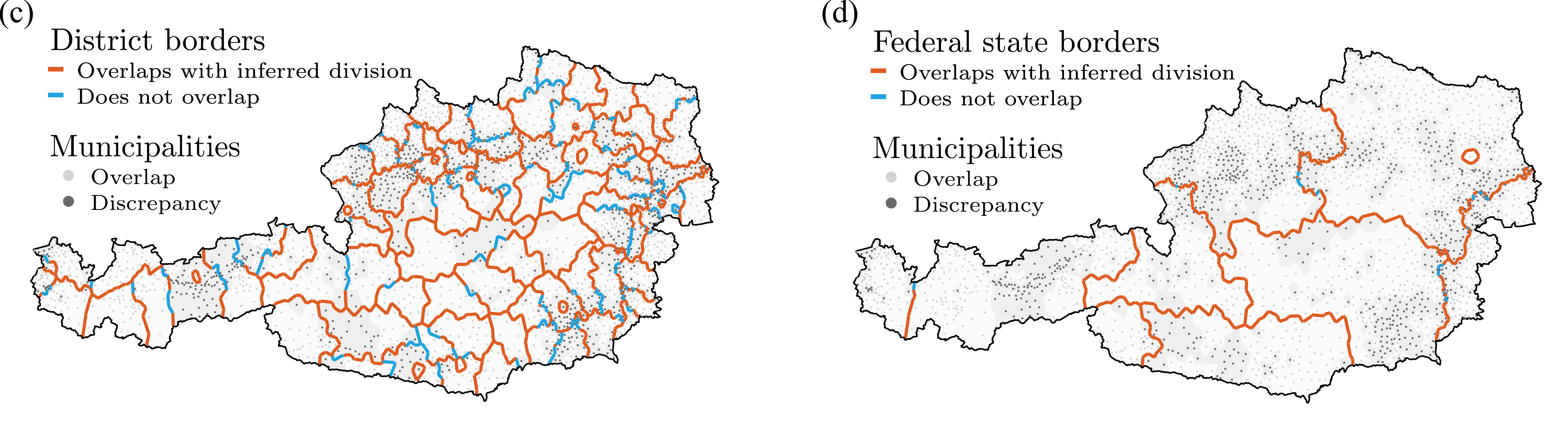}
  
  \caption{Comparison between administrative borders and the boundaries between
    groups of the municipalities inferred with the WSBM for 2013. If an
    administrative border coincides with the inferred partition, it is shown in
    red, otherwise in blue. Dark gray nodes indicate areas of discrepancy
    between the districts and the inferred partition at level $l=0$. Panel (a)
    shows how the borders between districts are recovered by the inferred
    divisions, corresponding to a recall of $0.47$, and (b) with the federal
    states, where the recall increases to $0.72$. Panels (c) and (d) show the
    same as (a) and (b), respectively, but for a binarized network, where an
    edge either exists or not between two municipalities, depending on whether
    the migration count is nonzero. In this case, the recall becomes $0.78$ and
    $0.95$ for (c) and (d), respectively.}
    \label{fig:boundary}
\end{figure}

\begin{figure}
 \centering
  \begin{overpic}[width=0.4\linewidth]{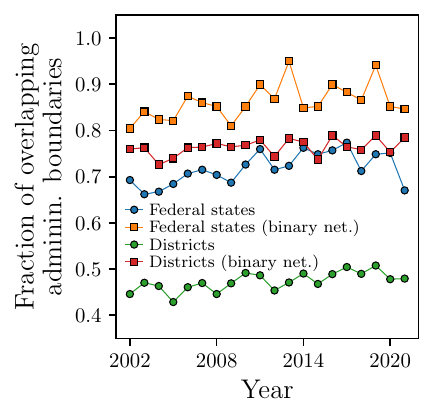}\put(-1,100){(a)}\end{overpic}
  \begin{overpic}[width=0.4\linewidth]{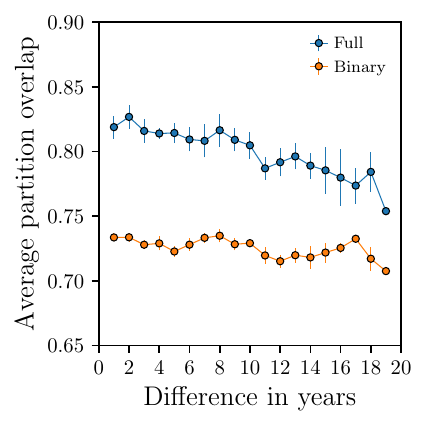}\put(-1,100){(b)}\end{overpic}
  \caption{(a) Recall of the administrative boundaries, with respect to the
    inferred partition, over a span of 20 years, for both districts and federal
    states, considering both the full network and the binarized version, as
    shown in the legend. (b) Average overlap between pairs of inferred
    partitions, as a function of the distance between them in years, both for
    the full network, as well as the binarized version.}
  \label{fig:boundary-long}
\end{figure}

What is perhaps the most striking feature of this inferred regionalization is
how well it aligns with administrative boundaries. In Fig.~\ref{fig:boundary}a
and b we show the overlap between districts and federal states with the
boundaries between the inferred groups, which show a remarkable agreement:
Around $45\%$ of the district boundaries coincide exactly with the boundaries
between the inferred groups, and the same happens for around $72\%$ of the
federal state boundaries. The overlap becomes even more pronounced once we
perform the same inference for a binarized version of the network, where we
ignore the magnitude of migrations, and consider instead a simple directed
graph, where an edge exists between two municipalities if any non-zero number of
migrations have occurred between them. In this case, the percentage of
administrative regions that match the inferred divisions increases to $78\%$ and
$95\%$, for district and federal state boundaries, respectively. (We emphasize
that our inferred clustering is based only on the number of migrations between
municipalities, without including any direct geographical information of any
kind.) This observation might be counterintuitive, as one could expect that
incorporating migration magnitudes would reinforce existing patterns. Instead,
including the magnitudes seems to diminish the visibility of district-level
effects. This may be because the most substantial migration flows occur between
broader regions. This would thereby reduce the relative influence of districts compared to larger administrative divisions such as NUTS1 (general areas) and NUTS2
(federal states), when the migration magnitudes are considered. This result
highlights that while administrative boundaries do seem to act as barriers to
migration, their influence is not uniform and varies depending on the intensity
of the flows. The substantial match between inferred and administrative
boundaries is not an exception for the year 2013, and persists instead for the
whole 20 years studied, as shown in Fig.~\ref{fig:boundary-long}a.
This stability is due to the inferred partitions showing only a small variation over time, as can be seen in the 
comparison in Fig.~\ref{fig:boundary-long}b, which reveals a very high partition overlap~\cite{peixoto_revealing_2021} between different times, with
only a very tenuous decay for longer time differences.

\begin{figure}
  \centering
     \includegraphics[width=\linewidth]{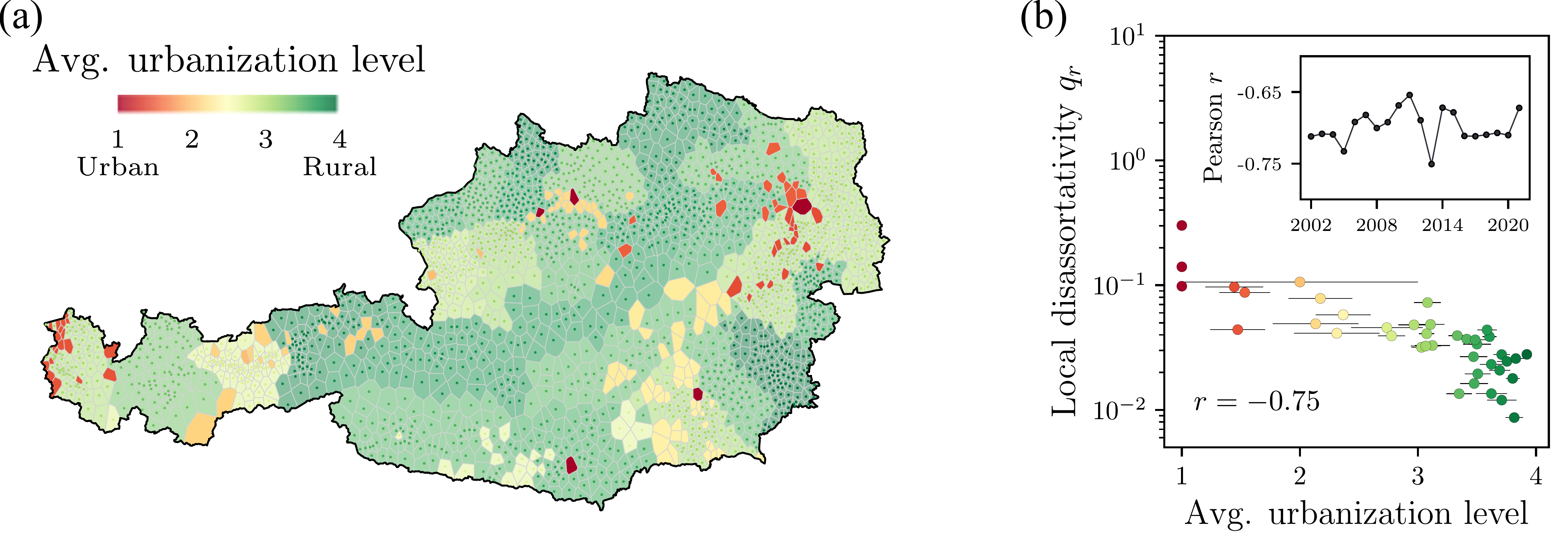}
  \caption{(a) Average urbanization level within the inferred communities at
    hierarchical level $l=0$, based on the urban-rural classification from
    Statistik Austria. (b) Negative correlation between local disassortativity
    of the communities (as defined in the text) and their urbanization level.
    Error bars represent the standard error of the urban-rural typology within
    each inferred group. The inset shows the robustness of this pattern across
    different years.}
  \label{fig:urban}
\end{figure}

In addition to the
overlap with the boundaries, and groups which are geographically localized, we
also observe subdivisions of the administrative districts, as well as groups
that are spatially discontiguous and composed of municipalities that are
geographically distant. Interestingly, these fragmented communities tend to
include more urbanized areas such as large cities. We show this in
Fig.~\ref{fig:urban}, which contains the urbanization level of the inferred
groups, defined as the urban-rural scores assigned to its constituent
municipalities, based on Statistik Austria’s classification~\cite{migration_data}: 1 corresponds to
urban centers, 2 to regional centers, 3 to rural areas surrounding centers, and
4 to rural areas. To further quantify how this relates to our inferred
clustering, we calculate the local disassortativity $q_r$ of group $r$, defined
as the ratio of a group's off-diagonal interactions to the total off-diagonal
entries in the group affinity matrix $\bm{m}=\{m_{rs}\}$, where $m_{rs}$ is the
number of migrations from group $s$ to group $r$,
\begin{equation}
\label{eq:local_disassortativity}
    q_r = \frac{\sum_{s: s\neq r} (m_{rs}+m_{sr})}{\sum_{r,s: r\neq s}m_{rs}}.
\end{equation}
This quantity serves as a proxy of how central a group is, with higher values
indicating stronger affinity with a wider variety of groups. Notably, groups
with more urbanization (i.e., lower urban-rural score) exhibit greater local
disassortativity, suggesting they function as hubs in the migration network. In
contrast, rural communities tend to be more self-contained. We quantify this
relationship using the Pearson correlation coefficient, which for 2013 is
$r = -0.75$, indicating a strong negative correlation between urban-rural level
and local disassortativity. These results are consistent across the period from
2002 to 2021, as shown in Fig.~\ref{fig:urban}b.

\begin{figure}[t]
    \centering
    \includegraphics[width=\linewidth]{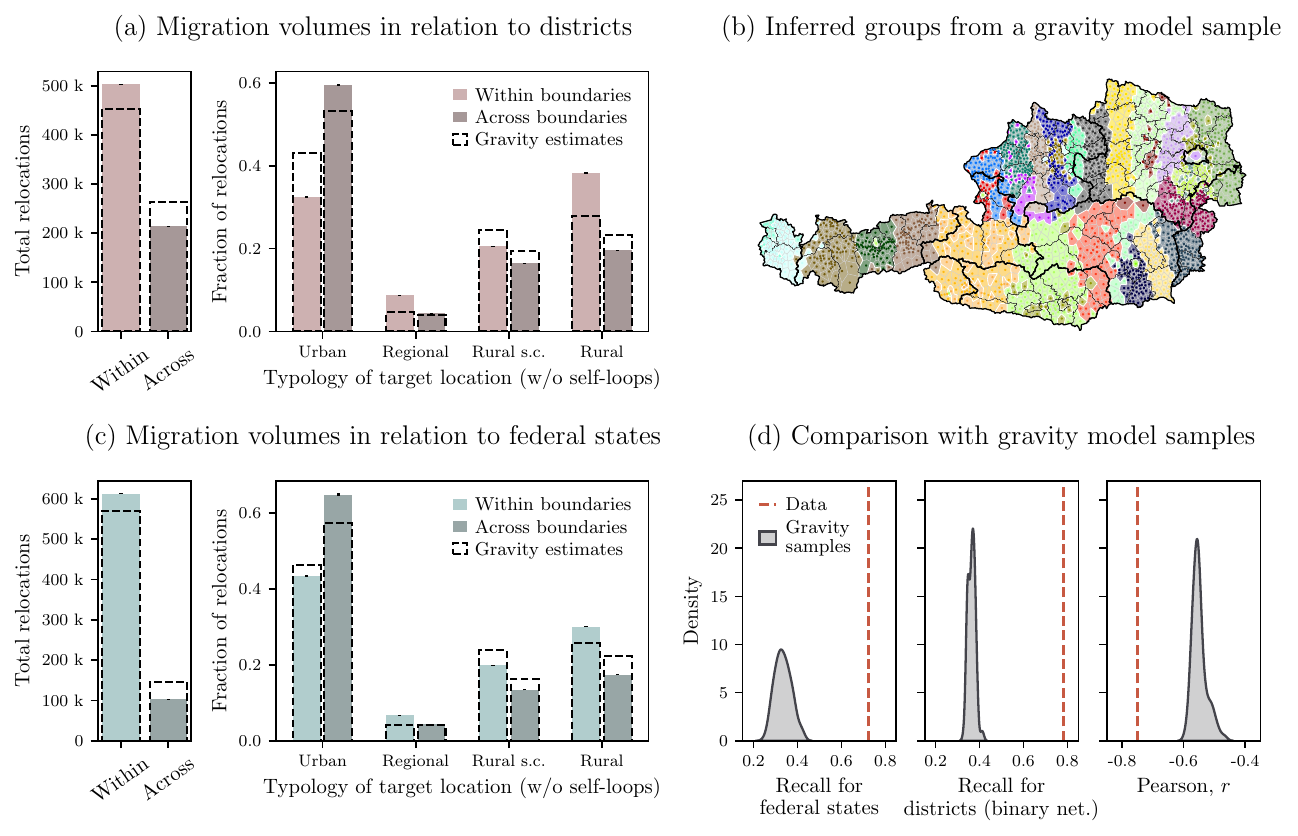}
    \caption{(a) Left: total number of migration events within and across
      district borders in 2013. Right: fraction of classification of the target
      location for within and across district boundaries migration events. The
      dashed bars are the corresponding quantities predicted by the gravity
      model. (c) Same as above for boundaries between federal states. (b) Groups
      inferred with the WSBM at hierarchical level $l=0$ on artificial data
      sampled from the best-fitting gravity model. (d) Comparison between
      different descriptors in the data and in samples from the gravity model.
      Left:
      recall of the federal state boundaries from the inferred partitions.
      Center: recall of the district boundaries from the inferred partitions
      with in the binarized network. Right: Pearson correlation between the
      local disassortativity and the urbanization level.}
    \label{fig:fig3}
\end{figure}

The observed interaction patterns differ notably from those predicted by the
gravity model, as shown in more detail in Figs.~\ref{fig:fig3}a and c, where we
compare the observed number of migrations across and within administrative
boundaries, with the respective expected number according to the best-fitting
gravity model. Likewise, in the same figure we do the same comparison with
migrations across and inside urban and rural regions. Relative to the gravity
model, we observe a stronger tendency for individuals to move within
administrative boundaries and a lower frequency of cross-border movements.
Significant deviations also emerge when considering the urban–rural
classification of the target location. Within administrative boundaries (at the
district or federal state level), individuals are more likely to relocate to
rural centers and less to urban centers than predicted by the gravity model.
Conversely, when crossing administrative borders, there is a pronounced
preference for moving to and from urban areas, well above model expectations.

To further detail the discrepancies with the gravity model, we generate 100
synthetic migration data for each year using the best fitting parameters for the
data, and apply the WSBM analysis to these artificial samples, as we did with
the empirical data. Fig.~\ref{fig:fig3}c shows the inferred groups at
hierarchical level $l=0$ for one of the samples, which differ significantly from
those identified in the empirical data. In particular, in the synthetic data,
the inferred group structures appear to be driven primarily by distance and population
effects, showing little alignment with Austria's regionalization or
administrative borders, which in this case become merely incidental. This
observation is reinforced by comparing the recall of federal states and
districts from the boundaries of the inferred groups (Fig.~\ref{fig:fig3}d). A
similar deviation is observed with respect to the urban-rural divide: the
negative correlation between local disassortativity and urbanization level found
in the empirical data is much stronger than in the synthetic samples. Taken
together, these results show that the gravity model produces samples that fail
to replicate key structures and dynamics observed in the real data, highlighting
the importance of network inference methods that are more agnostic, and hence
are able to uncover meaningful patterns in migration data.

\section{Discussion}

We have analyzed registry-level data covering all internal migrations in Austria
for a period of 20 years, using an inferential clustering methodology that
contrasts with the typical approaches used in this context. Such traditional analyses often focus on
regression of migration flows with covariates, flow frequency as a function of
distance and population density. In contrast, our approach fits an open-ended
generative model to the data, which is capable of uncovering statistically
significant mixing patterns that do not need to be specifically posited \emph{a
  priori.} Our method revealed substantial deviations from the gravity model
ansatz often used to characterize mobility data, consisting on strong
regionalization, a pervasive imprint of administrative boundaries, and an
exacerbated urban-rural divide.

The strong correlation of our inferential clustering with administrative
boundaries is particularly noteworthy. Similar overlaps with internal
administrative boundaries have been observed in other types of human behavior,
such as telephone calls in the UK~\cite{ratti2010redrawing}, online social
networks in the Netherlands~\cite{menyhert2025connectivity}, geo-tagging
trajectories in social media in China~\cite{liu2014uncovering} and
UK~\cite{yin_depicting_2017}, bank note circulation in the
US~\cite{thiemann_structure_2010}, daily commutes in
Italy~\cite{de2013commuter}, and mobile device tracking in
China~\cite{wei_urban_2024} and Hungary~\cite{pinter_quantifying_2024}. For
internal migration in particular, the situation is different: although other
studies have performed descriptive clustering analyses for
Austria~\cite{pitoski_network_2021}, Croatia~\cite{pitoski_network_2021-1},
Turkey~\cite{gursoy_investigating_2022} and Italy~\cite{sarra_network_2025}, and
have also found evidence of regionalization, the precise overlap with
administrative boundaries had not yet been quantified.

It is generally expected that administrative boundaries can reflect differences
in institutions, public services, and overall economic
development~\cite{jeong_discontinuities_2024}. In the case of commuting and
daily mobility behavior, as well as migration within the same municipality, it
is easy to imagine how administrative demarcations like official zoning, public
transport prices, location of schools and public services, and proximity to
commercial and/or industrial areas will have a significant impact on how people
move or relocate. These movements can then influence local socioeconomic
development, real estate prices, crime rates, etc., which in turn can cause
changes in the demarcation themselves, and so on. However, in the case of
relocation between municipalities and federal states, the precise mechanism that
makes administrative boundaries relevant is less clear. Importantly, our
analysis is not able to determine whether the boundaries act as \emph{causal}
barriers for migration, or if the district delineation simply captures the
natural ways people would migrate inside the country even if the boundaries
would not exist. On the one hand, these internal boundaries have a long history,
and capture demographic, cultural, and socio-economical structures of the
population that have persisted for a long time. Although the kind of formal
municipalities (``Gemeinde'') most comparable to today's Austria can be traced
back to 1849~\cite{noauthor_onb-alex_nodate}, their \emph{de facto} origin lies
often in the middle ages. On the other hand, from the second half of the 20th
century, several federal states implemented municipal reforms, often resulting
in the merger of municipalities (the latest finalized in the state of Styria in
2015\footnote{\url{https://www.ris.bka.gv.at/Dokumente/LgblAuth/LGBLA_ST_20140402_31/LGBLA_ST_20140402_31.pdfsig}},
which reduced its number of municipalities from 542 to 287). Some of these
merges were reversed decades later, since the local inhabitants felt they did
not adequately match the population structures. The actual division of the
municipalities into districts (``Bezirke'') stems from 1868, but likewise
reflects much older structures, and has also been subject to several reforms in
the 20th and 21st centuries, although most of these were rather minor and
localized. Our results show that in the 20-year period we analyze, the overlap
of our inferred clusters with the administrative boundaries is fairly stable,
indicating that the regionalization of internal migration is not deviating
significantly.

Regionalization of internal migration had also been previously observed in
Austria~\cite{pitoski_network_2021}, and other countries mentioned previously~\cite{pitoski_network_2021-1, gursoy_investigating_2022, sarra_network_2025}, and
high urban-rural flows have also been identified in the
US~\cite{xu2018structure} and Brazil~\cite{carvalho2020evolution}. However, both
of these phenomena can to some extent be associated with the gravity ansatz,
since nearby regions are expected to experience higher migration flows, as well
as to and from urban centers, which are highly populated. What our analysis
shows is that the gravity ansatz coupled with the population distribution of
Austria is not sufficient to account for the observed regionalization of
migration, and the movements between rural and urban centers.

The residuals of our inferred model with respect to the gravity ansatz revealed
a heterogeneous picture, where migrations between regions are either in excess
or deficit following a particular pattern. It would be of interest for a future
study to correlate such deviations with demographic and socioeconomic
indicators, to reveal potential mechanistic explanations. Such explanation could
potentially be used to forecast migration, and inform governmental policy.

\section{Data}\label{sec:data}

In this study, we use the publicly available version of the \textsc{MIGSTAT - Wanderungsstatistik} dataset~\cite{migration_data}, which was collected by Statistik Austria, the Federal Statistical Office of Austria. 
This dataset contains all relocations within Austria that involve a change in main residence under registration law, for the period 2002-2021.
We consider each of these change of main residence as a migration event.
The data are aggregated at the municipality level (i.e., $N=2093$), and we analyze each year separately, excluding the information about gender and nationality associated with the relocations.

Our analysis also incorporates the yearly population data (collected at the beginning of each year) of all Austrian municipalities, and the urban-rural typology of these municipalities.
Both datasets are provided by Statistik Austria.
The urban-rural classification assigns an integer value in the interval $[1,4]$ to each municipality, with the following mapping: $1\to$ urban center, $2\to$ regional center, $3\to$ rural area surrounding center, and $4\to$ rural area.
This assignment is produced by integrating information regarding the administrative status of municipalities, the population density across the geographical areas, as well as the infrastructure facilities and the presence of commuting relationships.

A summary of the dataset, with the total population $p_{\rm tot}$, number of edges $E$ in the network, and total number of migration events $M$ per year is shown in Table~\ref{tab:tab1}.

\begin{table}
    \centering
    \begin{adjustbox}{width=\columnwidth,center}
    \begin{tabular}{c*{10}{c}}
 & \textbf{2002} & \textbf{2003} & \textbf{2004} & \textbf{2005} & \textbf{2006} & \textbf{2007} & \textbf{2008} & \textbf{2009} & \textbf{2010} & \textbf{2011} \\ \hline \hline
$p_{\rm tot}$ & 8063640 & 8100273 & 8142573 & 8201359 & 8254298 & 8282984 & 8307989 & 8335003 & 8351643 & 8375164 \\
$E$ & 61829 & 62152 & 64194 & 65404 & 66150 & 67565 & 70203 & 70372 & 70830 & 72412 \\
$M$ & 649153 & 627078 & 650411 & 664616 & 672125 & 689167 & 692698 & 685115 & 681538 & 701242 \\
\hline

    \end{tabular}
    \end{adjustbox}

\vspace{1em}
\begin{adjustbox}{width=\columnwidth,center}
    \begin{tabular}{c*{10}{c}}
 & \textbf{2012} & \textbf{2013} & \textbf{2014} & \textbf{2015} & \textbf{2016} & \textbf{2017} & \textbf{2018} & \textbf{2019} & \textbf{2020} & \textbf{2021} \\ \hline \hline
$p_{\rm tot}$ & 8408121 & 8451860 & 8507786 & 8584926 & 8700471 & 8772865 & 8822267 & 8858775 & 8901064 & 8932664 \\
$E$ & 72555 & 73422 & 74126 & 78016 & 80733 & 78597 & 78476 & 78393 & 78496 & 78640 \\
$M$ & 714697 & 716436 & 739918 & 795028 & 817139 & 801624 & 797666 & 798420 & 781472 & 782995 \\ \hline
    \end{tabular}
\end{adjustbox}
    \caption{Total population, number of edges in the network, and total number of migration events for each year considered in the dataset. The number of nodes is constant across time and equal to 2093.}
    \label{tab:tab1}
\end{table}

\section{Methods}\label{sec:methods}
\subsection{Inferring gravity models of migration}
Gravity models are among the most widely used tools for analyzing migration and
mobility phenomena. The key idea behind these models is that when choosing to
migrate, people face the interplay between the cost and the opportunity
associated with moving. Typically, the cost of migration is modeled as a
function of the distance between the origin and destination of a migration
event, while the opportunity is represented as a function of the product of the
population sizes of the two locations. More sophisticated versions of gravity
models incorporate additional variables, such as economic or socio-demographic
indicators, to better capture the attractiveness of a location or to refine the
estimation of moving costs. However, including these additional factors
increases the complexity of model and its inference. In this study, we focus on
the basic form of the gravity model to allow for clearer interpretation and to
draw conclusions from the general formulation, rather than relying on \emph{ad hoc} specifications. In the simple formulation outlined above, the expected
amount of migrations $x_{ij}$ from a location $j$ to a location $i$ with population
$p_j$ and $p_i$ respectively, and separated by distance $d_{ij}$, can be
expressed as
\begin{equation}
\label{eq:gravity_traditional}
\mu_{ij}=K\frac{\left(p_ip_j\right)^{\alpha}}{d_{ij}^{\beta}},
\end{equation}
where $\alpha$ and $\beta$ are non-negative real parameters controlling the effect on the population product and distance attributes, while $K$---also a non-negative real parameter---adjusts the overall magnitude of migration flows.

A common practice in applied research is to take the logarithm of both sides of Eq.~\eqref{eq:gravity_traditional} and estimate the parameters $K$, $\alpha$, and $\beta$ by finding the best fit of the resulting functional relational form.
However, this approach introduces several misspecifications~\cite{burger2009specification}, including the treatment of zero values and the biases caused by the logarithmic transformation.
Moreover, it overlooks the fact that $\mu_{ij}$ in Eq.~\eqref{eq:gravity_traditional} is a positive real number, while our empirical observations are count values, i.e., non-negative integers.
These limitations can be overcome by modeling migration flows between two locations as independent Poisson-distributed random variables, with the expected value (or rate) defined by the functional form of the gravity law~\cite{flowerdew1982method}, i.e.
\begin{equation}
\label{eq:gravity_pois}
    P\left(\bm{x}|K,\alpha,\beta\right)=\prod_{i,j} \frac{\mu_{ij}^{x_{ij}}\mathrm{e}^{-\mu_{ij}}}{x_{ij}!}.
\end{equation}
To apply the gravity model to the Austrian migration data analyzed in this
study, we need to address the presence of intra-municipality migrations, i.e.,
self-loops in the network with $d_{ij}=0$. For that, we introduce a specific
formulation for the Poisson rate in the case where $i=j$,
\begin{equation}
\label{eq:gravity_pois_self}
    \mu_{ij} = 
    \begin{cases}
        K\frac{(p_ip_j)^{\alpha}}{d_{ij}^{\beta}} & \text{if } i\neq j \\
        Cp_i^{\delta} & \text{if } i = j
    \end{cases}
\end{equation}
where $C$ and $\delta$ are non-negative real parameters independently capturing
intra-municipality migration dynamics. Based on this formulation, the model
parameters ($K, \alpha, \beta, C, \delta$) are inferred from data on migration counts $\bm{x}$ according to the joint posterior distribution
\begin{equation}
  P(K, \alpha, \beta, C, \delta \mid \bm{x}) = \frac{P(\bm{x} \mid K, \alpha, \beta, C, \delta)P(K, \alpha, \beta, C, \delta)}{P(\bm{x})},
\end{equation}
with $P(K, \alpha, \beta, C, \delta) \propto 1$ being a noninformative prior. Samples from this posterior distribution are obtained using Hamiltonian Monte Carlo~\cite{betancourt2013hamiltonianmontecarlohierarchical,  gelman2013bayesian} as implemented in Stan~\cite{carpenter_stan_2017}, which allows us to quantify the parameter uncertainties.



\subsection{Non-parametric weighted stochastic block models}
In our analysis, we employ a class of stochastic block models (SBMs) where
weights are represented as edge covariates~\cite{peixoto2018nonparametric}.
These are generative models for networks that, in addition to the adjacency
matrix $\bm{A}=\{A_{ij}\}$, also have real or discrete edge covariates
$\bm{x} = \{x_{ij}\}$. This formulation decouples the edge existence from the
presence of a null weight, i.e., the non-existence of an edge is different from
an edge with zero weight. In the migration network under study we have a binary,
asymmetric adjacency matrix, with $A_{ij}=1$ if there is at least one migration
from location $j$ to location $i$ and $A_{ij}=0$ otherwise, and on each existing
edge we have one positive integer weight $x_{ij}\in \mathbb{N}^{+}$ representing
the total number of relocation events from location $j$ to location $i$. The
underlying assumption of the SBM is that nodes are divided into $B$ groups, with
each node having a value $b_i\in \{0,...,B-1\}$ specifying its group membership.
In addition to the edge placement, the edge weights are also sampled according
to the group membership of the source and target node. This means that both the
edges $\bm{A}$ and the edge weights $\bm{x}$ are sampled from parametric
distributions conditioned on the group memberships of the nodes, i.e.
\begin{equation}
\label{eq:wSBM_edges_weights}
    P\left(\bm{A},\bm{x}|\bm{\theta},\bm{\gamma}, \bm{b}\right) =
    P\left(\bm{x}|\bm{A}, \bm{\gamma}, \bm{b}\right)
    P\left(\bm{A}| \bm{\theta}, \bm{b}\right)
\end{equation}
with the weights being sampled conditioned on the existence of the edges,
\begin{equation}
\label{eq:wSBM_weights}
P\left(\bm{x}|\bm{A}, \bm{\gamma}, \bm{b}\right)=\prod_{rs} P\left(x_{rs}|\gamma_{rs}\right)
\end{equation}
where $x_{rs}=\left\{x_{ij}|A_{ij}>0 \; \wedge \; (b_i,b_j)=(r,s)\right\}$ are the covariates between groups $r$ and $s$, with $\gamma_{rs}$ being the parameters controlling the sampling of the weights, specific to the group pair.
The placement of the edges is independent of the edge covariates and controlled by the parameters $\bm{\theta}$.

Given the generative model presented above---with a specific choice for the edge placements and the weight distributions---the set of parameters $\bm{\theta}$ and $\bm{\gamma}$  could be estimated via maximum likelihood.
Doing so would lead to overfitting as the likelihood would increase with the complexity of the model.
A more principled way to proceed is to obtain the Bayesian posterior probability for the partitions $\bm{b}$, i.e.
\begin{equation}
\label{eq:bayes_groups}
    P\left(\bm{b}|\bm{A},\bm{x}\right)=\frac{P\left(\bm{A},\bm{x}|\bm{b}\right)P(\bm{b})}{P\left(\bm{A},\bm{x}\right)},
\end{equation}
where the numerator is the marginal likelihood of the observed network and weights, integrated over the model parameters,\begin{equation}
\label{eq:marginal_likelihood}
    P\left(\bm{A},\bm{x}|\bm{b}\right)=
    \sum_{\theta} \int P\left(\bm{A},\bm{x}|\bm{\theta},\bm{\gamma}, \bm{b}\right)
    P(\bm{\theta})P(\bm{\gamma})\text{d}\bm{\gamma}=
    P(\bm{A}|\bm{b})P(\bm{x}|\bm{A},\bm{b}).
\end{equation}
For the edge placement---i.e. corresponding to $P(\bm{A}|\bm{b})P(\bm{b})$---we
use the nested microcanonical degree-corrected SBM described in
Ref.~\cite{peixoto2014hierarchical}. The degree
correction~\cite{karrer_stochastic_2011} of the model prescribes, in addition to
the modular group structure, that the networks generated by the model possess a
specific out- and in-degree sequence, $\bm{k}=\{k_i^{+},k_i^{-}\}$. The edges are
placed according to
\begin{equation}
\label{eq:wSBM_edges_DC}
    P\left(\bm{A}|\bm{\theta}=\{\bm{e},\bm{k}\},\bm{b}\right)=
    \frac{\prod_{i}k_{i}^{+}!k_{i}^{-}!\prod_{rs} e_{rs}!}{\prod_{r}e_{r}^{+}!e_{r}^{-}!\prod_{ij} A_{ij}!}
\end{equation}
where $\bm{e}=\{e_{rs}\}$ specifies the number of edges that are placed between
groups. The nested nature of this model is given by considering, as part of the
prior probabilities for the edge placements, a hierarchy of multigraphs. The
groups in the observed network are considered themselves as nodes of a smaller
multigraph that is also generated by the SBM, with its nodes put in their own
groups, forming an even smaller multigraph, and so on recursively. This gives a
nested hierarchy with $L$ levels $\{\bm{b}^l\}=\{\{b_r^{(l)}\}_l\}$, where
$b_r^{(l)} \in \{0,...,B_l-1\}$ is the group membership of the group/node $r$ at
level $l\in\{0,...,L-1\}$, with the boundary condition that at the topmost level
there is only one group, $B_{L}=1$. The adjacency of the multigraph at level
$l$ is
\begin{equation}
\label{eq:wSBM_adjacency}
    e_{rs}^l = \sum_{t,u} e_{tu}^{l-1} \delta_{b_t^{(l)},r} \delta_{b_u^{(l)},s}
\end{equation}
with $e_{ij}^{0}=A_{ij}$ as a boundary condition. This hierarchy of partitions
allows for a multilevel description of the network data under study---describing
in the migration network the mesoscopic structure of the migration flow at
different geographical resolutions.

The marginal likelihood of the edge weights completes the nonparametric approach presented in Ref.~\cite{peixoto2018nonparametric}.
This is obtained by integrating over the weight parameters $\bm{\gamma}$,
\begin{equation}
\label{eq:weights_marginal}
    P\left(\bm{x}|\bm{A},\bm{b}\right)=
    \int P\left(\bm{x}|\bm{A},\bm{\gamma}, \bm{b}\right) P(\bm{\gamma})\text{d}\bm{\gamma} =
    \prod_{rs} \int P\left(x_{rs}|\gamma_{rs}\right)P(\gamma_{rs})\text{d}\gamma_{rs}.
\end{equation}
This formulation of the model allows us to consider a variety of elementary
choices for the terms appearing in Eq.~\eqref{eq:weights_marginal}, reflecting
the nature of the covariates (e.g. continuous or discrete, signed or not,
bounded or unbounded). Among the different formulations compatible with the
characteristics of the covariates under study, we select the best model
according to the choice that yields the smallest description length---the
negative log-likelihood for models with discrete covariates---or by computing
the posterior odds ratio for continuous models, as described in
Refs.~\cite{peixoto2017nonparametric, peixoto2018nonparametric}. For the
networks of Austrian migrations, where edges weights correspond to relocation
counts, i.e. $x_{ij}\in \mathbb{N}^{+}$, the two elementary choices are to
consider a Poisson or a geometric distribution. The model with weights sampled
from geometric distributions consistently provided a better fit (i.e. smaller
description length) for the data under study, which is compatible with the
underdispersion of Poisson formulations found previously in the
literature~\cite{burger2009specification}.

Given a choice for the marginal likelihood in the numerator of
Eq.~\eqref{eq:bayes_groups}, we find the best hierarchical partition of the
network under study by employing the efficient agglomerative multilevel Markov
chain Monte Carlo (MCMC) algorithm described in
Refs.~\cite{peixoto2014hierarchical, peixoto2014efficient} and further refining
the results running the MCMC at null temperature. This procedure is performed
for 10 different random initializations, where after running the agglomerative
heuristic, we run the greedy MCMC for $3 \times 10 ^ 4$ sweeps. Out of these 10
states, we select the one yielding the lowest description length.

\section*{Acknowledgments}
The authors acknowledge support from the project ``MOMA: Multiscale network 
modeling of migration flows in Austria'' funded by WWTF (Grant ID: 10.47379/ESS22032). 
The computational results have been achieved using the Austrian Scientific Computing (ASC) infrastructure. We also thank A. Malek and M. Czaika for fruitful discussion on this work.

\bibliography{references,references2}

\begin{thebibliography}{58}%
\makeatletter
\providecommand \@ifxundefined [1]{%
 \@ifx{#1\undefined}
}%
\providecommand \@ifnum [1]{%
 \ifnum #1\expandafter \@firstoftwo
 \else \expandafter \@secondoftwo
 \fi
}%
\providecommand \@ifx [1]{%
 \ifx #1\expandafter \@firstoftwo
 \else \expandafter \@secondoftwo
 \fi
}%
\providecommand \natexlab [1]{#1}%
\providecommand \enquote  [1]{``#1''}%
\providecommand \bibnamefont  [1]{#1}%
\providecommand \bibfnamefont [1]{#1}%
\providecommand \citenamefont [1]{#1}%
\providecommand \href@noop [0]{\@secondoftwo}%
\providecommand \href [0]{\begingroup \@sanitize@url \@href}%
\providecommand \@href[1]{\@@startlink{#1}\@@href}%
\providecommand \@@href[1]{\endgroup#1\@@endlink}%
\providecommand \@sanitize@url [0]{\catcode `\\12\catcode `\$12\catcode
  `\&12\catcode `\#12\catcode `\^12\catcode `\_12\catcode `\%12\relax}%
\providecommand \@@startlink[1]{}%
\providecommand \@@endlink[0]{}%
\providecommand \url  [0]{\begingroup\@sanitize@url \@url }%
\providecommand \@url [1]{\endgroup\@href {#1}{\urlprefix }}%
\providecommand \urlprefix  [0]{URL }%
\providecommand \Eprint [0]{\href }%
\providecommand \doibase [0]{http://dx.doi.org/}%
\providecommand \selectlanguage [0]{\@gobble}%
\providecommand \bibinfo  [0]{\@secondoftwo}%
\providecommand \bibfield  [0]{\@secondoftwo}%
\providecommand \translation [1]{[#1]}%
\providecommand \BibitemOpen [0]{}%
\providecommand \bibitemStop [0]{}%
\providecommand \bibitemNoStop [0]{.\EOS\space}%
\providecommand \EOS [0]{\spacefactor3000\relax}%
\providecommand \BibitemShut  [1]{\csname bibitem#1\endcsname}%
\let\auto@bib@innerbib\@empty
\bibitem [{\citenamefont {Papademetriou}\ and\ \citenamefont
  {Martin}(1991)}]{papademetriou1991unsettled}%
  \BibitemOpen
  \bibfield  {author} {\bibinfo {author} {\bibfnamefont {D.~G.}\ \bibnamefont
  {Papademetriou}}\ and\ \bibinfo {author} {\bibfnamefont {P.~L.}\ \bibnamefont
  {Martin}},\ }\href@noop {} {\emph {\bibinfo {title} {The unsettled
  relationship: Labor migration and economic development}}},\ \bibinfo {number}
  {33}\ (\bibinfo  {publisher} {Greenwood Publishing Group},\ \bibinfo {year}
  {1991})\BibitemShut {NoStop}%
\bibitem [{\citenamefont {Czaika}\ and\ \citenamefont
  {Reinprecht}(2022)}]{czaika2022migration}%
  \BibitemOpen
  \bibfield  {author} {\bibinfo {author} {\bibfnamefont {M.}~\bibnamefont
  {Czaika}}\ and\ \bibinfo {author} {\bibfnamefont {C.}~\bibnamefont
  {Reinprecht}},\ }in\ \href {\doibase
  https://doi.org/10.1007/978-3-030-92377-8_3} {\emph {\bibinfo {booktitle}
  {Introduction to migration studies: An interactive guide to the literatures
  on migration and diversity}}}\ (\bibinfo  {publisher} {Springer International
  Publishing Cham},\ \bibinfo {year} {2022})\ pp.\ \bibinfo {pages}
  {49--82}\BibitemShut {NoStop}%
\bibitem [{\citenamefont {Scholten}\ \emph {et~al.}(2022)\citenamefont
  {Scholten}, \citenamefont {Pisarevskaya},\ and\ \citenamefont
  {Levy}}]{scholten2022introduction}%
  \BibitemOpen
  \bibfield  {author} {\bibinfo {author} {\bibfnamefont {P.}~\bibnamefont
  {Scholten}}, \bibinfo {author} {\bibfnamefont {A.}~\bibnamefont
  {Pisarevskaya}}, \ and\ \bibinfo {author} {\bibfnamefont {N.}~\bibnamefont
  {Levy}},\ }in\ \href {\doibase https://doi.org/10.1007/978-3-030-92377-8_1}
  {\emph {\bibinfo {booktitle} {Introduction to Migration Studies: An
  Interactive Guide to the Literatures on Migration and Diversity}}}\ (\bibinfo
   {publisher} {Springer},\ \bibinfo {year} {2022})\ pp.\ \bibinfo {pages}
  {3--24}\BibitemShut {NoStop}%
\bibitem [{\citenamefont {Barbosa}\ \emph {et~al.}(2018)\citenamefont
  {Barbosa}, \citenamefont {Barthelemy}, \citenamefont {Ghoshal}, \citenamefont
  {James}, \citenamefont {Lenormand}, \citenamefont {Louail}, \citenamefont
  {Menezes}, \citenamefont {Ramasco}, \citenamefont {Simini},\ and\
  \citenamefont {Tomasini}}]{barbosa2018human}%
  \BibitemOpen
  \bibfield  {author} {\bibinfo {author} {\bibfnamefont {H.}~\bibnamefont
  {Barbosa}}, \bibinfo {author} {\bibfnamefont {M.}~\bibnamefont {Barthelemy}},
  \bibinfo {author} {\bibfnamefont {G.}~\bibnamefont {Ghoshal}}, \bibinfo
  {author} {\bibfnamefont {C.~R.}\ \bibnamefont {James}}, \bibinfo {author}
  {\bibfnamefont {M.}~\bibnamefont {Lenormand}}, \bibinfo {author}
  {\bibfnamefont {T.}~\bibnamefont {Louail}}, \bibinfo {author} {\bibfnamefont
  {R.}~\bibnamefont {Menezes}}, \bibinfo {author} {\bibfnamefont {J.~J.}\
  \bibnamefont {Ramasco}}, \bibinfo {author} {\bibfnamefont {F.}~\bibnamefont
  {Simini}}, \ and\ \bibinfo {author} {\bibfnamefont {M.}~\bibnamefont
  {Tomasini}},\ }\href {\doibase https://doi.org/10.1016/j.physrep.2018.01.001}
  {\bibfield  {journal} {\bibinfo  {journal} {Physics Reports}\ }\textbf
  {\bibinfo {volume} {734}},\ \bibinfo {pages} {1} (\bibinfo {year} {2018})},\
  \bibinfo {note} {human mobility: Models and applications}\BibitemShut
  {NoStop}%
\bibitem [{\citenamefont {Brockmann}\ \emph {et~al.}(2006)\citenamefont
  {Brockmann}, \citenamefont {Hufnagel},\ and\ \citenamefont
  {Geisel}}]{brockmann2006scaling}%
  \BibitemOpen
  \bibfield  {author} {\bibinfo {author} {\bibfnamefont {D.}~\bibnamefont
  {Brockmann}}, \bibinfo {author} {\bibfnamefont {L.}~\bibnamefont {Hufnagel}},
  \ and\ \bibinfo {author} {\bibfnamefont {T.}~\bibnamefont {Geisel}},\ }\href
  {\doibase 10.1038/nature04292} {\bibfield  {journal} {\bibinfo  {journal}
  {Nature}\ }\textbf {\bibinfo {volume} {439}},\ \bibinfo {pages} {462}
  (\bibinfo {year} {2006})}\BibitemShut {NoStop}%
\bibitem [{\citenamefont {Gonzalez}\ \emph {et~al.}(2008)\citenamefont
  {Gonzalez}, \citenamefont {Hidalgo},\ and\ \citenamefont
  {Barabasi}}]{gonzalez2008understanding}%
  \BibitemOpen
  \bibfield  {author} {\bibinfo {author} {\bibfnamefont {M.~C.}\ \bibnamefont
  {Gonzalez}}, \bibinfo {author} {\bibfnamefont {C.~A.}\ \bibnamefont
  {Hidalgo}}, \ and\ \bibinfo {author} {\bibfnamefont {A.-L.}\ \bibnamefont
  {Barabasi}},\ }\href {\doibase 10.1038/nature06958} {\bibfield  {journal}
  {\bibinfo  {journal} {nature}\ }\textbf {\bibinfo {volume} {453}},\ \bibinfo
  {pages} {779} (\bibinfo {year} {2008})}\BibitemShut {NoStop}%
\bibitem [{\citenamefont {Song}\ \emph {et~al.}(2010)\citenamefont {Song},
  \citenamefont {Koren}, \citenamefont {Wang},\ and\ \citenamefont
  {Barab{\'a}si}}]{song2010modelling}%
  \BibitemOpen
  \bibfield  {author} {\bibinfo {author} {\bibfnamefont {C.}~\bibnamefont
  {Song}}, \bibinfo {author} {\bibfnamefont {T.}~\bibnamefont {Koren}},
  \bibinfo {author} {\bibfnamefont {P.}~\bibnamefont {Wang}}, \ and\ \bibinfo
  {author} {\bibfnamefont {A.-L.}\ \bibnamefont {Barab{\'a}si}},\ }\href
  {\doibase 10.1038/nphys1760} {\bibfield  {journal} {\bibinfo  {journal}
  {Nature physics}\ }\textbf {\bibinfo {volume} {6}},\ \bibinfo {pages} {818}
  (\bibinfo {year} {2010})}\BibitemShut {NoStop}%
\bibitem [{\citenamefont {Zipf}(1946)}]{zipf1946the}%
  \BibitemOpen
  \bibfield  {author} {\bibinfo {author} {\bibfnamefont {G.~K.}\ \bibnamefont
  {Zipf}},\ }\href {http://www.jstor.org/stable/2087063} {\bibfield  {journal}
  {\bibinfo  {journal} {American Sociological Review}\ }\textbf {\bibinfo
  {volume} {11}},\ \bibinfo {pages} {677} (\bibinfo {year} {1946})}\BibitemShut
  {NoStop}%
\bibitem [{\citenamefont {Lewer}\ and\ \citenamefont {Van~den
  Berg}(2008)}]{lewer2008gravity}%
  \BibitemOpen
  \bibfield  {author} {\bibinfo {author} {\bibfnamefont {J.~J.}\ \bibnamefont
  {Lewer}}\ and\ \bibinfo {author} {\bibfnamefont {H.}~\bibnamefont {Van~den
  Berg}},\ }\href {\doibase 10.1016/j.econlet.2007.06.019} {\bibfield
  {journal} {\bibinfo  {journal} {Economics letters}\ }\textbf {\bibinfo
  {volume} {99}},\ \bibinfo {pages} {164} (\bibinfo {year} {2008})}\BibitemShut
  {NoStop}%
\bibitem [{\citenamefont {Prieto~Curiel}\ \emph {et~al.}(2018)\citenamefont
  {Prieto~Curiel}, \citenamefont {Pappalardo}, \citenamefont {Gabrielli},\ and\
  \citenamefont {Bishop}}]{prieto2018gravity}%
  \BibitemOpen
  \bibfield  {author} {\bibinfo {author} {\bibfnamefont {R.}~\bibnamefont
  {Prieto~Curiel}}, \bibinfo {author} {\bibfnamefont {L.}~\bibnamefont
  {Pappalardo}}, \bibinfo {author} {\bibfnamefont {L.}~\bibnamefont
  {Gabrielli}}, \ and\ \bibinfo {author} {\bibfnamefont {S.~R.}\ \bibnamefont
  {Bishop}},\ }\href {\doibase 10.1371/journal.pone.0199892} {\bibfield
  {journal} {\bibinfo  {journal} {PloS one}\ }\textbf {\bibinfo {volume}
  {13}},\ \bibinfo {pages} {e0199892} (\bibinfo {year} {2018})}\BibitemShut
  {NoStop}%
\bibitem [{\citenamefont {Cabanas-Tirapu}\ \emph {et~al.}(2025)\citenamefont
  {Cabanas-Tirapu}, \citenamefont {Dan{\'u}s}, \citenamefont {Moro},
  \citenamefont {Sales-Pardo},\ and\ \citenamefont
  {Guimer{\`a}}}]{cabanas2025human}%
  \BibitemOpen
  \bibfield  {author} {\bibinfo {author} {\bibfnamefont {O.}~\bibnamefont
  {Cabanas-Tirapu}}, \bibinfo {author} {\bibfnamefont {L.}~\bibnamefont
  {Dan{\'u}s}}, \bibinfo {author} {\bibfnamefont {E.}~\bibnamefont {Moro}},
  \bibinfo {author} {\bibfnamefont {M.}~\bibnamefont {Sales-Pardo}}, \ and\
  \bibinfo {author} {\bibfnamefont {R.}~\bibnamefont {Guimer{\`a}}},\ }\href
  {\doibase 10.1038/s41467-025-56495-5} {\bibfield  {journal} {\bibinfo
  {journal} {Nature Communications}\ }\textbf {\bibinfo {volume} {16}},\
  \bibinfo {pages} {1336} (\bibinfo {year} {2025})}\BibitemShut {NoStop}%
\bibitem [{\citenamefont {Mazzoli}\ \emph {et~al.}(2019)\citenamefont
  {Mazzoli}, \citenamefont {Molas}, \citenamefont {Bassolas}, \citenamefont
  {Lenormand}, \citenamefont {Colet},\ and\ \citenamefont
  {Ramasco}}]{mazzoli2019field}%
  \BibitemOpen
  \bibfield  {author} {\bibinfo {author} {\bibfnamefont {M.}~\bibnamefont
  {Mazzoli}}, \bibinfo {author} {\bibfnamefont {A.}~\bibnamefont {Molas}},
  \bibinfo {author} {\bibfnamefont {A.}~\bibnamefont {Bassolas}}, \bibinfo
  {author} {\bibfnamefont {M.}~\bibnamefont {Lenormand}}, \bibinfo {author}
  {\bibfnamefont {P.}~\bibnamefont {Colet}}, \ and\ \bibinfo {author}
  {\bibfnamefont {J.~J.}\ \bibnamefont {Ramasco}},\ }\href@noop {} {\bibfield
  {journal} {\bibinfo  {journal} {Nature communications}\ }\textbf {\bibinfo
  {volume} {10}},\ \bibinfo {pages} {3895} (\bibinfo {year}
  {2019})}\BibitemShut {NoStop}%
\bibitem [{\citenamefont {Kwon}\ \emph {et~al.}(2023)\citenamefont {Kwon},
  \citenamefont {Hong}, \citenamefont {Jung},\ and\ \citenamefont
  {Jo}}]{kwon2023multiple}%
  \BibitemOpen
  \bibfield  {author} {\bibinfo {author} {\bibfnamefont {O.-H.}\ \bibnamefont
  {Kwon}}, \bibinfo {author} {\bibfnamefont {I.}~\bibnamefont {Hong}}, \bibinfo
  {author} {\bibfnamefont {W.-S.}\ \bibnamefont {Jung}}, \ and\ \bibinfo
  {author} {\bibfnamefont {H.-H.}\ \bibnamefont {Jo}},\ }\href@noop {}
  {\bibfield  {journal} {\bibinfo  {journal} {EPJ Data Science}\ }\textbf
  {\bibinfo {volume} {12}},\ \bibinfo {pages} {57} (\bibinfo {year}
  {2023})}\BibitemShut {NoStop}%
\bibitem [{\citenamefont {Liu}\ \emph {et~al.}(2014)\citenamefont {Liu},
  \citenamefont {Sui}, \citenamefont {Kang},\ and\ \citenamefont
  {Gao}}]{liu2014uncovering}%
  \BibitemOpen
  \bibfield  {author} {\bibinfo {author} {\bibfnamefont {Y.}~\bibnamefont
  {Liu}}, \bibinfo {author} {\bibfnamefont {Z.}~\bibnamefont {Sui}}, \bibinfo
  {author} {\bibfnamefont {C.}~\bibnamefont {Kang}}, \ and\ \bibinfo {author}
  {\bibfnamefont {Y.}~\bibnamefont {Gao}},\ }\href@noop {} {\bibfield
  {journal} {\bibinfo  {journal} {PloS one}\ }\textbf {\bibinfo {volume} {9}},\
  \bibinfo {pages} {e86026} (\bibinfo {year} {2014})}\BibitemShut {NoStop}%
\bibitem [{\citenamefont {Li}\ \emph {et~al.}(2021)\citenamefont {Li},
  \citenamefont {Gao}, \citenamefont {Luo}, \citenamefont {Yao}, \citenamefont
  {Chen}, \citenamefont {Shang}, \citenamefont {Jiang},\ and\ \citenamefont
  {Stanley}}]{li2021gravity}%
  \BibitemOpen
  \bibfield  {author} {\bibinfo {author} {\bibfnamefont {R.}~\bibnamefont
  {Li}}, \bibinfo {author} {\bibfnamefont {S.}~\bibnamefont {Gao}}, \bibinfo
  {author} {\bibfnamefont {A.}~\bibnamefont {Luo}}, \bibinfo {author}
  {\bibfnamefont {Q.}~\bibnamefont {Yao}}, \bibinfo {author} {\bibfnamefont
  {B.}~\bibnamefont {Chen}}, \bibinfo {author} {\bibfnamefont {F.}~\bibnamefont
  {Shang}}, \bibinfo {author} {\bibfnamefont {R.}~\bibnamefont {Jiang}}, \ and\
  \bibinfo {author} {\bibfnamefont {H.~E.}\ \bibnamefont {Stanley}},\
  }\href@noop {} {\bibfield  {journal} {\bibinfo  {journal} {Physical Review
  E}\ }\textbf {\bibinfo {volume} {103}},\ \bibinfo {pages} {012312} (\bibinfo
  {year} {2021})}\BibitemShut {NoStop}%
\bibitem [{\citenamefont {Expert}\ \emph {et~al.}(2011)\citenamefont {Expert},
  \citenamefont {Evans}, \citenamefont {Blondel},\ and\ \citenamefont
  {Lambiotte}}]{expert2011uncovering}%
  \BibitemOpen
  \bibfield  {author} {\bibinfo {author} {\bibfnamefont {P.}~\bibnamefont
  {Expert}}, \bibinfo {author} {\bibfnamefont {T.~S.}\ \bibnamefont {Evans}},
  \bibinfo {author} {\bibfnamefont {V.~D.}\ \bibnamefont {Blondel}}, \ and\
  \bibinfo {author} {\bibfnamefont {R.}~\bibnamefont {Lambiotte}},\ }\href
  {\doibase 10.1073/pnas.1018962108} {\bibfield  {journal} {\bibinfo  {journal}
  {Proceedings of the National Academy of Sciences}\ }\textbf {\bibinfo
  {volume} {108}},\ \bibinfo {pages} {7663} (\bibinfo {year}
  {2011})}\BibitemShut {NoStop}%
\bibitem [{\citenamefont {Poot}\ \emph {et~al.}(2016)\citenamefont {Poot},
  \citenamefont {Alimi}, \citenamefont {Cameron},\ and\ \citenamefont
  {Mar{\'e}}}]{poot2016gravity}%
  \BibitemOpen
  \bibfield  {author} {\bibinfo {author} {\bibfnamefont {J.}~\bibnamefont
  {Poot}}, \bibinfo {author} {\bibfnamefont {O.}~\bibnamefont {Alimi}},
  \bibinfo {author} {\bibfnamefont {M.~P.}\ \bibnamefont {Cameron}}, \ and\
  \bibinfo {author} {\bibfnamefont {D.~C.}\ \bibnamefont {Mar{\'e}}},\
  }\href@noop {} {\bibfield  {journal} {\bibinfo  {journal} {Investigaciones
  Regionales-Journal of Regional Research}\ ,\ \bibinfo {pages} {63}} (\bibinfo
  {year} {2016})}\BibitemShut {NoStop}%
\bibitem [{\citenamefont {Stouffer}(1940)}]{stouffer1940intervening}%
  \BibitemOpen
  \bibfield  {author} {\bibinfo {author} {\bibfnamefont {S.~A.}\ \bibnamefont
  {Stouffer}},\ }\href {http://www.jstor.org/stable/2084520} {\bibfield
  {journal} {\bibinfo  {journal} {American Sociological Review}\ }\textbf
  {\bibinfo {volume} {5}},\ \bibinfo {pages} {845} (\bibinfo {year}
  {1940})}\BibitemShut {NoStop}%
\bibitem [{\citenamefont {Akwawua}\ and\ \citenamefont
  {Pooler}(2000)}]{akwawua2000intervening}%
  \BibitemOpen
  \bibfield  {author} {\bibinfo {author} {\bibfnamefont {S.}~\bibnamefont
  {Akwawua}}\ and\ \bibinfo {author} {\bibfnamefont {J.}~\bibnamefont
  {Pooler}},\ }in\ \href {https://grf.bgu.ac.il/index.php/GRF/article/view/214}
  {\emph {\bibinfo {booktitle} {Geography Research Forum}}}\ (\bibinfo
  {organization} {Ben-Gurion University of the Negev Press},\ \bibinfo {year}
  {2000})\ pp.\ \bibinfo {pages} {33--51}\BibitemShut {NoStop}%
\bibitem [{\citenamefont {Simini}\ \emph {et~al.}(2012)\citenamefont {Simini},
  \citenamefont {Gonz{\'a}lez}, \citenamefont {Maritan},\ and\ \citenamefont
  {Barab{\'a}si}}]{simini2012universal}%
  \BibitemOpen
  \bibfield  {author} {\bibinfo {author} {\bibfnamefont {F.}~\bibnamefont
  {Simini}}, \bibinfo {author} {\bibfnamefont {M.~C.}\ \bibnamefont
  {Gonz{\'a}lez}}, \bibinfo {author} {\bibfnamefont {A.}~\bibnamefont
  {Maritan}}, \ and\ \bibinfo {author} {\bibfnamefont {A.-L.}\ \bibnamefont
  {Barab{\'a}si}},\ }\href {\doibase 10.1038/nature10856} {\bibfield  {journal}
  {\bibinfo  {journal} {Nature}\ }\textbf {\bibinfo {volume} {484}},\ \bibinfo
  {pages} {96} (\bibinfo {year} {2012})}\BibitemShut {NoStop}%
\bibitem [{\citenamefont {Lenormand}\ \emph {et~al.}(2016)\citenamefont
  {Lenormand}, \citenamefont {Bassolas},\ and\ \citenamefont
  {Ramasco}}]{lenormand2016systematic}%
  \BibitemOpen
  \bibfield  {author} {\bibinfo {author} {\bibfnamefont {M.}~\bibnamefont
  {Lenormand}}, \bibinfo {author} {\bibfnamefont {A.}~\bibnamefont {Bassolas}},
  \ and\ \bibinfo {author} {\bibfnamefont {J.~J.}\ \bibnamefont {Ramasco}},\
  }\href@noop {} {\bibfield  {journal} {\bibinfo  {journal} {Journal of
  Transport Geography}\ }\textbf {\bibinfo {volume} {51}},\ \bibinfo {pages}
  {158} (\bibinfo {year} {2016})}\BibitemShut {NoStop}%
\bibitem [{\citenamefont {Yang}\ \emph {et~al.}(2014)\citenamefont {Yang},
  \citenamefont {Herrera}, \citenamefont {Eagle},\ and\ \citenamefont
  {Gonz{\'a}lez}}]{yang2014limits}%
  \BibitemOpen
  \bibfield  {author} {\bibinfo {author} {\bibfnamefont {Y.}~\bibnamefont
  {Yang}}, \bibinfo {author} {\bibfnamefont {C.}~\bibnamefont {Herrera}},
  \bibinfo {author} {\bibfnamefont {N.}~\bibnamefont {Eagle}}, \ and\ \bibinfo
  {author} {\bibfnamefont {M.~C.}\ \bibnamefont {Gonz{\'a}lez}},\ }\href@noop
  {} {\bibfield  {journal} {\bibinfo  {journal} {Scientific reports}\ }\textbf
  {\bibinfo {volume} {4}},\ \bibinfo {pages} {5662} (\bibinfo {year}
  {2014})}\BibitemShut {NoStop}%
\bibitem [{\citenamefont {Burger}\ \emph {et~al.}(2009)\citenamefont {Burger},
  \citenamefont {Van~Oort},\ and\ \citenamefont
  {Linders}}]{burger2009specification}%
  \BibitemOpen
  \bibfield  {author} {\bibinfo {author} {\bibfnamefont {M.}~\bibnamefont
  {Burger}}, \bibinfo {author} {\bibfnamefont {F.}~\bibnamefont {Van~Oort}}, \
  and\ \bibinfo {author} {\bibfnamefont {G.-J.}\ \bibnamefont {Linders}},\
  }\href {\doibase 10.1080/17421770902834327} {\bibfield  {journal} {\bibinfo
  {journal} {Spatial economic analysis}\ }\textbf {\bibinfo {volume} {4}},\
  \bibinfo {pages} {167} (\bibinfo {year} {2009})}\BibitemShut {NoStop}%
\bibitem [{\citenamefont {Masucci}\ \emph {et~al.}(2013)\citenamefont
  {Masucci}, \citenamefont {Serras}, \citenamefont {Johansson},\ and\
  \citenamefont {Batty}}]{masucci2013gravity}%
  \BibitemOpen
  \bibfield  {author} {\bibinfo {author} {\bibfnamefont {A.~P.}\ \bibnamefont
  {Masucci}}, \bibinfo {author} {\bibfnamefont {J.}~\bibnamefont {Serras}},
  \bibinfo {author} {\bibfnamefont {A.}~\bibnamefont {Johansson}}, \ and\
  \bibinfo {author} {\bibfnamefont {M.}~\bibnamefont {Batty}},\ }\href@noop {}
  {\bibfield  {journal} {\bibinfo  {journal} {Physical Review E—Statistical,
  Nonlinear, and Soft Matter Physics}\ }\textbf {\bibinfo {volume} {88}},\
  \bibinfo {pages} {022812} (\bibinfo {year} {2013})}\BibitemShut {NoStop}%
\bibitem [{\citenamefont {Beyer}\ \emph {et~al.}(2022)\citenamefont {Beyer},
  \citenamefont {Schewe},\ and\ \citenamefont
  {Lotze-Campen}}]{beyer2022gravity}%
  \BibitemOpen
  \bibfield  {author} {\bibinfo {author} {\bibfnamefont {R.~M.}\ \bibnamefont
  {Beyer}}, \bibinfo {author} {\bibfnamefont {J.}~\bibnamefont {Schewe}}, \
  and\ \bibinfo {author} {\bibfnamefont {H.}~\bibnamefont {Lotze-Campen}},\
  }\href@noop {} {\bibfield  {journal} {\bibinfo  {journal} {Humanities and
  Social Sciences Communications}\ }\textbf {\bibinfo {volume} {9}},\ \bibinfo
  {pages} {1} (\bibinfo {year} {2022})}\BibitemShut {NoStop}%
\bibitem [{\citenamefont {Peixoto}(2019)}]{peixoto_bayesian_2019}%
  \BibitemOpen
  \bibfield  {author} {\bibinfo {author} {\bibfnamefont {T.~P.}\ \bibnamefont
  {Peixoto}},\ }in\ \href {\doibase 10.1002/9781119483298.ch11} {\emph
  {\bibinfo {booktitle} {Advances in {{Network Clustering}} and
  {{Blockmodeling}}}}}\ (\bibinfo  {publisher} {John Wiley \& Sons, Ltd},\
  \bibinfo {year} {2019})\ pp.\ \bibinfo {pages} {289--332},\ \Eprint
  {http://arxiv.org/abs/1705.10225} {arXiv:1705.10225} \BibitemShut {NoStop}%
\bibitem [{\citenamefont {Peixoto}(2018)}]{peixoto2018nonparametric}%
  \BibitemOpen
  \bibfield  {author} {\bibinfo {author} {\bibfnamefont {T.~P.}\ \bibnamefont
  {Peixoto}},\ }\href {\doibase 10.1103/PhysRevE.97.012306} {\bibfield
  {journal} {\bibinfo  {journal} {Physical Review E}\ }\textbf {\bibinfo
  {volume} {97}},\ \bibinfo {pages} {012306} (\bibinfo {year}
  {2018})}\BibitemShut {NoStop}%
\bibitem [{\citenamefont {Grünwald}(2007)}]{grunwald2007minimum}%
  \BibitemOpen
  \bibfield  {author} {\bibinfo {author} {\bibfnamefont {P.~D.}\ \bibnamefont
  {Grünwald}},\ }\href@noop {} {\emph {\bibinfo {title} {The Minimum
  Description Length Principle}}}\ (\bibinfo  {publisher} {The MIT Press},\
  \bibinfo {address} {Cambridge, MA},\ \bibinfo {year} {2007})\BibitemShut
  {NoStop}%
\bibitem [{\citenamefont {Rissanen}(2010)}]{rissanen2010information}%
  \BibitemOpen
  \bibfield  {author} {\bibinfo {author} {\bibfnamefont {J.}~\bibnamefont
  {Rissanen}},\ }\href@noop {} {\emph {\bibinfo {title} {Information and
  Complexity in Statistical Modeling}}},\ \bibinfo {edition} {1st}\ ed.\
  (\bibinfo  {publisher} {Springer},\ \bibinfo {address} {New York},\ \bibinfo
  {year} {2010})\BibitemShut {NoStop}%
\bibitem [{\citenamefont
  {Peixoto}(2014{\natexlab{a}})}]{peixoto_hierarchical_2014}%
  \BibitemOpen
  \bibfield  {author} {\bibinfo {author} {\bibfnamefont {T.~P.}\ \bibnamefont
  {Peixoto}},\ }\href {\doibase 10.1103/PhysRevX.4.011047} {\bibfield
  {journal} {\bibinfo  {journal} {Physical Review X}\ }\textbf {\bibinfo
  {volume} {4}},\ \bibinfo {pages} {011047} (\bibinfo {year}
  {2014}{\natexlab{a}})},\ \Eprint {http://arxiv.org/abs/1310.4377}
  {arXiv:1310.4377} \BibitemShut {NoStop}%
\bibitem [{\citenamefont {Betancourt}\ and\ \citenamefont
  {Girolami}(2013)}]{betancourt2013hamiltonianmontecarlohierarchical}%
  \BibitemOpen
  \bibfield  {author} {\bibinfo {author} {\bibfnamefont {M.~J.}\ \bibnamefont
  {Betancourt}}\ and\ \bibinfo {author} {\bibfnamefont {M.}~\bibnamefont
  {Girolami}},\ }\href {https://arxiv.org/abs/1312.0906} {\enquote {\bibinfo
  {title} {Hamiltonian monte carlo for hierarchical models},}\ } (\bibinfo
  {year} {2013}),\ \Eprint {http://arxiv.org/abs/1312.0906} {arXiv:1312.0906
  [stat.ME]} \BibitemShut {NoStop}%
\bibitem [{\citenamefont {Gelman}\ \emph {et~al.}(2013)\citenamefont {Gelman},
  \citenamefont {Carlin}, \citenamefont {Stern}, \citenamefont {Dunson},
  \citenamefont {Vehtari},\ and\ \citenamefont {Rubin}}]{gelman2013bayesian}%
  \BibitemOpen
  \bibfield  {author} {\bibinfo {author} {\bibfnamefont {A.}~\bibnamefont
  {Gelman}}, \bibinfo {author} {\bibfnamefont {J.~B.}\ \bibnamefont {Carlin}},
  \bibinfo {author} {\bibfnamefont {H.~S.}\ \bibnamefont {Stern}}, \bibinfo
  {author} {\bibfnamefont {D.~B.}\ \bibnamefont {Dunson}}, \bibinfo {author}
  {\bibfnamefont {A.}~\bibnamefont {Vehtari}}, \ and\ \bibinfo {author}
  {\bibfnamefont {D.~B.}\ \bibnamefont {Rubin}},\ }\enquote {\bibinfo {title}
  {Bayesian data analysis},}\ \ (\bibinfo  {publisher} {Chapman and Hall/CRC},\
  \bibinfo {year} {2013})\ Chap.~\bibinfo {chapter} {12}\BibitemShut {NoStop}%
\bibitem [{\citenamefont {Peixoto}(2023)}]{peixoto_descriptive_2023}%
  \BibitemOpen
  \bibfield  {author} {\bibinfo {author} {\bibfnamefont {T.~P.}\ \bibnamefont
  {Peixoto}},\ }\href {\doibase 10.1017/9781009118897} {\bibfield  {journal}
  {\bibinfo  {journal} {Elements in the Structure and Dynamics of Complex
  Networks}\ } (\bibinfo {year} {2023}),\ 10.1017/9781009118897},\ \Eprint
  {http://arxiv.org/abs/2112.00183} {arXiv:2112.00183} \BibitemShut {NoStop}%
\bibitem [{\citenamefont {Zhang}\ and\ \citenamefont
  {Peixoto}(2020)}]{zhang_statistical_2020}%
  \BibitemOpen
  \bibfield  {author} {\bibinfo {author} {\bibfnamefont {L.}~\bibnamefont
  {Zhang}}\ and\ \bibinfo {author} {\bibfnamefont {T.~P.}\ \bibnamefont
  {Peixoto}},\ }\href {\doibase 10.1103/PhysRevResearch.2.043271} {\bibfield
  {journal} {\bibinfo  {journal} {Physical Review Research}\ }\textbf {\bibinfo
  {volume} {2}},\ \bibinfo {pages} {043271} (\bibinfo {year} {2020})},\ \Eprint
  {http://arxiv.org/abs/2006.14493} {arXiv:2006.14493} \BibitemShut {NoStop}%
\bibitem [{\citenamefont {Holten}(2006)}]{holten2006hierarchical}%
  \BibitemOpen
  \bibfield  {author} {\bibinfo {author} {\bibfnamefont {D.}~\bibnamefont
  {Holten}},\ }\href {\doibase 10.1109/TVCG.2006.147} {\bibfield  {journal}
  {\bibinfo  {journal} {IEEE Transactions on Visualization and Computer
  Graphics}\ }\textbf {\bibinfo {volume} {12}},\ \bibinfo {pages} {741}
  (\bibinfo {year} {2006})}\BibitemShut {NoStop}%
\bibitem [{\citenamefont {Peixoto}(2021)}]{peixoto_revealing_2021}%
  \BibitemOpen
  \bibfield  {author} {\bibinfo {author} {\bibfnamefont {T.~P.}\ \bibnamefont
  {Peixoto}},\ }\href {\doibase 10.1103/PhysRevX.11.021003} {\bibfield
  {journal} {\bibinfo  {journal} {Physical Review X}\ }\textbf {\bibinfo
  {volume} {11}},\ \bibinfo {pages} {021003} (\bibinfo {year} {2021})},\
  \Eprint {http://arxiv.org/abs/2005.13977} {arXiv:2005.13977} \BibitemShut
  {NoStop}%
\bibitem [{mig()}]{migration_data}%
  \BibitemOpen
  \href@noop {} {}\bibinfo {howpublished} {Available at
  \url{https://data.statistik.gv.at/}}\BibitemShut {NoStop}%
\bibitem [{\citenamefont {Ratti}\ \emph {et~al.}(2010)\citenamefont {Ratti},
  \citenamefont {Sobolevsky}, \citenamefont {Calabrese}, \citenamefont
  {Andris}, \citenamefont {Reades}, \citenamefont {Martino}, \citenamefont
  {Claxton},\ and\ \citenamefont {Strogatz}}]{ratti2010redrawing}%
  \BibitemOpen
  \bibfield  {author} {\bibinfo {author} {\bibfnamefont {C.}~\bibnamefont
  {Ratti}}, \bibinfo {author} {\bibfnamefont {S.}~\bibnamefont {Sobolevsky}},
  \bibinfo {author} {\bibfnamefont {F.}~\bibnamefont {Calabrese}}, \bibinfo
  {author} {\bibfnamefont {C.}~\bibnamefont {Andris}}, \bibinfo {author}
  {\bibfnamefont {J.}~\bibnamefont {Reades}}, \bibinfo {author} {\bibfnamefont
  {M.}~\bibnamefont {Martino}}, \bibinfo {author} {\bibfnamefont
  {R.}~\bibnamefont {Claxton}}, \ and\ \bibinfo {author} {\bibfnamefont
  {S.~H.}\ \bibnamefont {Strogatz}},\ }\href@noop {} {\bibfield  {journal}
  {\bibinfo  {journal} {PloS one}\ }\textbf {\bibinfo {volume} {5}},\ \bibinfo
  {pages} {e14248} (\bibinfo {year} {2010})}\BibitemShut {NoStop}%
\bibitem [{\citenamefont {Menyh{\'e}rt}\ \emph {et~al.}(2025)\citenamefont
  {Menyh{\'e}rt}, \citenamefont {Bok{\'a}nyi}, \citenamefont {Corten},
  \citenamefont {Heemskerk}, \citenamefont {Kazmina},\ and\ \citenamefont
  {Takes}}]{menyhert2025connectivity}%
  \BibitemOpen
  \bibfield  {author} {\bibinfo {author} {\bibfnamefont {M.}~\bibnamefont
  {Menyh{\'e}rt}}, \bibinfo {author} {\bibfnamefont {E.}~\bibnamefont
  {Bok{\'a}nyi}}, \bibinfo {author} {\bibfnamefont {R.}~\bibnamefont {Corten}},
  \bibinfo {author} {\bibfnamefont {E.~M.}\ \bibnamefont {Heemskerk}}, \bibinfo
  {author} {\bibfnamefont {Y.}~\bibnamefont {Kazmina}}, \ and\ \bibinfo
  {author} {\bibfnamefont {F.~W.}\ \bibnamefont {Takes}},\ }\href@noop {}
  {\bibfield  {journal} {\bibinfo  {journal} {EPJ Data Science}\ }\textbf
  {\bibinfo {volume} {14}},\ \bibinfo {pages} {8} (\bibinfo {year}
  {2025})}\BibitemShut {NoStop}%
\bibitem [{\citenamefont {Yin}\ \emph {et~al.}(2017)\citenamefont {Yin},
  \citenamefont {~}, \citenamefont {~},\ and\ \citenamefont {{and
  Wang}}}]{yin_depicting_2017}%
  \BibitemOpen
  \bibfield  {author} {\bibinfo {author} {\bibfnamefont {J.}~\bibnamefont
  {Yin}}, \bibinfo {author} {\bibfnamefont {S.}~\bibnamefont {~}, \bibfnamefont
  {Aiman}}, \bibinfo {author} {\bibfnamefont {Y.}~\bibnamefont {~},
  \bibfnamefont {Dandong}}, \ and\ \bibinfo {author} {\bibfnamefont
  {S.}~\bibnamefont {{and Wang}}},\ }\href {\doibase
  10.1080/13658816.2017.1282615} {\bibfield  {journal} {\bibinfo  {journal}
  {International Journal of Geographical Information Science}\ }\textbf
  {\bibinfo {volume} {31}},\ \bibinfo {pages} {1293} (\bibinfo {year}
  {2017})}\BibitemShut {NoStop}%
\bibitem [{\citenamefont {Thiemann}\ \emph {et~al.}(2010)\citenamefont
  {Thiemann}, \citenamefont {Theis}, \citenamefont {Grady}, \citenamefont
  {Brune},\ and\ \citenamefont {Brockmann}}]{thiemann_structure_2010}%
  \BibitemOpen
  \bibfield  {author} {\bibinfo {author} {\bibfnamefont {C.}~\bibnamefont
  {Thiemann}}, \bibinfo {author} {\bibfnamefont {F.}~\bibnamefont {Theis}},
  \bibinfo {author} {\bibfnamefont {D.}~\bibnamefont {Grady}}, \bibinfo
  {author} {\bibfnamefont {R.}~\bibnamefont {Brune}}, \ and\ \bibinfo {author}
  {\bibfnamefont {D.}~\bibnamefont {Brockmann}},\ }\href {\doibase
  10.1371/journal.pone.0015422} {\bibfield  {journal} {\bibinfo  {journal}
  {PLOS ONE}\ }\textbf {\bibinfo {volume} {5}},\ \bibinfo {pages} {e15422}
  (\bibinfo {year} {2010})}\BibitemShut {NoStop}%
\bibitem [{\citenamefont {De~Montis}\ \emph {et~al.}(2013)\citenamefont
  {De~Montis}, \citenamefont {Caschili},\ and\ \citenamefont
  {Chessa}}]{de2013commuter}%
  \BibitemOpen
  \bibfield  {author} {\bibinfo {author} {\bibfnamefont {A.}~\bibnamefont
  {De~Montis}}, \bibinfo {author} {\bibfnamefont {S.}~\bibnamefont {Caschili}},
  \ and\ \bibinfo {author} {\bibfnamefont {A.}~\bibnamefont {Chessa}},\
  }\href@noop {} {\bibfield  {journal} {\bibinfo  {journal} {The European
  Physical Journal Special Topics}\ }\textbf {\bibinfo {volume} {215}},\
  \bibinfo {pages} {75} (\bibinfo {year} {2013})}\BibitemShut {NoStop}%
\bibitem [{\citenamefont {Wei}\ and\ \citenamefont
  {Zhao}(2024)}]{wei_urban_2024}%
  \BibitemOpen
  \bibfield  {author} {\bibinfo {author} {\bibfnamefont {Y.}~\bibnamefont
  {Wei}}\ and\ \bibinfo {author} {\bibfnamefont {X.}~\bibnamefont {Zhao}},\
  }\href {\doibase 10.1109/ACCESS.2024.3487921} {\bibfield  {journal} {\bibinfo
   {journal} {IEEE Access}\ }\textbf {\bibinfo {volume} {12}},\ \bibinfo
  {pages} {159635} (\bibinfo {year} {2024})}\BibitemShut {NoStop}%
\bibitem [{\citenamefont {Pint{\'e}r}\ and\ \citenamefont
  {Lengyel}(2024)}]{pinter_quantifying_2024}%
  \BibitemOpen
  \bibfield  {author} {\bibinfo {author} {\bibfnamefont {G.}~\bibnamefont
  {Pint{\'e}r}}\ and\ \bibinfo {author} {\bibfnamefont {B.}~\bibnamefont
  {Lengyel}},\ }\href {\doibase 10.48550/arXiv.2312.11343} {\enquote {\bibinfo
  {title} {Quantifying {{Barriers}} of {{Urban Mobility}}},}\ } (\bibinfo
  {year} {2024}),\ \Eprint {http://arxiv.org/abs/2312.11343} {arXiv:2312.11343
  [physics]} \BibitemShut {NoStop}%
\bibitem [{\citenamefont {Pitoski}\ \emph
  {et~al.}(2021{\natexlab{a}})\citenamefont {Pitoski}, \citenamefont
  {Lampoltshammer},\ and\ \citenamefont {Parycek}}]{pitoski_network_2021}%
  \BibitemOpen
  \bibfield  {author} {\bibinfo {author} {\bibfnamefont {D.}~\bibnamefont
  {Pitoski}}, \bibinfo {author} {\bibfnamefont {T.~J.}\ \bibnamefont
  {Lampoltshammer}}, \ and\ \bibinfo {author} {\bibfnamefont {P.}~\bibnamefont
  {Parycek}},\ }\href {\doibase 10.1145/3447539} {\bibfield  {journal}
  {\bibinfo  {journal} {Digital Government: Research and Practice}\ }\textbf
  {\bibinfo {volume} {2}},\ \bibinfo {pages} {25:1} (\bibinfo {year}
  {2021}{\natexlab{a}})}\BibitemShut {NoStop}%
\bibitem [{\citenamefont {Pitoski}\ \emph
  {et~al.}(2021{\natexlab{b}})\citenamefont {Pitoski}, \citenamefont
  {Lampoltshammer},\ and\ \citenamefont {Parycek}}]{pitoski_network_2021-1}%
  \BibitemOpen
  \bibfield  {author} {\bibinfo {author} {\bibfnamefont {D.}~\bibnamefont
  {Pitoski}}, \bibinfo {author} {\bibfnamefont {T.~J.}\ \bibnamefont
  {Lampoltshammer}}, \ and\ \bibinfo {author} {\bibfnamefont {P.}~\bibnamefont
  {Parycek}},\ }\href {\doibase 10.1186/s40649-021-00093-0} {\bibfield
  {journal} {\bibinfo  {journal} {Computational Social Networks}\ }\textbf
  {\bibinfo {volume} {8}},\ \bibinfo {pages} {1} (\bibinfo {year}
  {2021}{\natexlab{b}})}\BibitemShut {NoStop}%
\bibitem [{\citenamefont {G{\"u}rsoy}\ and\ \citenamefont
  {Badur}(2022)}]{gursoy_investigating_2022}%
  \BibitemOpen
  \bibfield  {author} {\bibinfo {author} {\bibfnamefont {F.}~\bibnamefont
  {G{\"u}rsoy}}\ and\ \bibinfo {author} {\bibfnamefont {B.}~\bibnamefont
  {Badur}},\ }\href {\doibase 10.1007/s13278-022-00974-w} {\bibfield  {journal}
  {\bibinfo  {journal} {Social Network Analysis and Mining}\ }\textbf {\bibinfo
  {volume} {12}},\ \bibinfo {pages} {150} (\bibinfo {year} {2022})}\BibitemShut
  {NoStop}%
\bibitem [{\citenamefont {Sarra}\ \emph {et~al.}(2025)\citenamefont {Sarra},
  \citenamefont {D'Ingiullo}, \citenamefont {Evangelista}, \citenamefont
  {Nissi}, \citenamefont {Quaglione},\ and\ \citenamefont
  {Di~Battista}}]{sarra_network_2025}%
  \BibitemOpen
  \bibfield  {author} {\bibinfo {author} {\bibfnamefont {A.}~\bibnamefont
  {Sarra}}, \bibinfo {author} {\bibfnamefont {D.}~\bibnamefont {D'Ingiullo}},
  \bibinfo {author} {\bibfnamefont {A.}~\bibnamefont {Evangelista}}, \bibinfo
  {author} {\bibfnamefont {E.}~\bibnamefont {Nissi}}, \bibinfo {author}
  {\bibfnamefont {D.}~\bibnamefont {Quaglione}}, \ and\ \bibinfo {author}
  {\bibfnamefont {T.}~\bibnamefont {Di~Battista}},\ }\href {\doibase
  10.1016/j.seps.2025.102225} {\bibfield  {journal} {\bibinfo  {journal}
  {Socio-Economic Planning Sciences}\ }\textbf {\bibinfo {volume} {100}},\
  \bibinfo {pages} {102225} (\bibinfo {year} {2025})}\BibitemShut {NoStop}%
\bibitem [{\citenamefont {Jeong}(2024)}]{jeong_discontinuities_2024}%
  \BibitemOpen
  \bibfield  {author} {\bibinfo {author} {\bibfnamefont {J.}~\bibnamefont
  {Jeong}},\ }\href {\doibase 10.1177/23998083231217013} {\bibfield  {journal}
  {\bibinfo  {journal} {Environment and Planning B: Urban Analytics and City
  Science}\ }\textbf {\bibinfo {volume} {51}},\ \bibinfo {pages} {1227}
  (\bibinfo {year} {2024})}\BibitemShut {NoStop}%
\bibitem [{noa()}]{noauthor_onb-alex_nodate}%
  \BibitemOpen
  \href
  {https://alex.onb.ac.at/cgi-content/alex?aid=rgb&datum=1849&size=45&page=339}
  {\enquote {\bibinfo {title} {{{{\"O}NB-ALEX}} - {{Reichsgesetzblatt}}
  1849-1918},}\ }\BibitemShut {NoStop}%
\bibitem [{\citenamefont {Xu}(2018)}]{xu2018structure}%
  \BibitemOpen
  \bibfield  {author} {\bibinfo {author} {\bibfnamefont {Z.}~\bibnamefont
  {Xu}},\ }\href@noop {} {\bibfield  {journal} {\bibinfo  {journal} {Papers in
  Regional Science}\ }\textbf {\bibinfo {volume} {97}},\ \bibinfo {pages} {785}
  (\bibinfo {year} {2018})}\BibitemShut {NoStop}%
\bibitem [{\citenamefont {Carvalho}\ and\ \citenamefont
  {Charles-Edwards}(2020)}]{carvalho2020evolution}%
  \BibitemOpen
  \bibfield  {author} {\bibinfo {author} {\bibfnamefont {R.~C.~d.}\
  \bibnamefont {Carvalho}}\ and\ \bibinfo {author} {\bibfnamefont
  {E.}~\bibnamefont {Charles-Edwards}},\ }\href@noop {} {\bibfield  {journal}
  {\bibinfo  {journal} {Population, Space and Place}\ }\textbf {\bibinfo
  {volume} {26}},\ \bibinfo {pages} {e2332} (\bibinfo {year}
  {2020})}\BibitemShut {NoStop}%
\bibitem [{\citenamefont {Flowerdew}\ and\ \citenamefont
  {Aitkin}(1982)}]{flowerdew1982method}%
  \BibitemOpen
  \bibfield  {author} {\bibinfo {author} {\bibfnamefont {R.}~\bibnamefont
  {Flowerdew}}\ and\ \bibinfo {author} {\bibfnamefont {M.}~\bibnamefont
  {Aitkin}},\ }\href {\doibase
  https://doi.org/10.1111/j.1467-9787.1982.tb00744.x} {\bibfield  {journal}
  {\bibinfo  {journal} {Journal of regional science}\ }\textbf {\bibinfo
  {volume} {22}} (\bibinfo {year} {1982}),\
  https://doi.org/10.1111/j.1467-9787.1982.tb00744.x}\BibitemShut {NoStop}%
\bibitem [{\citenamefont {Carpenter}\ \emph {et~al.}(2017)\citenamefont
  {Carpenter}, \citenamefont {Gelman}, \citenamefont {Hoffman}, \citenamefont
  {Lee}, \citenamefont {Goodrich}, \citenamefont {Betancourt}, \citenamefont
  {Brubaker}, \citenamefont {Guo}, \citenamefont {Li},\ and\ \citenamefont
  {Riddell}}]{carpenter_stan_2017}%
  \BibitemOpen
  \bibfield  {author} {\bibinfo {author} {\bibfnamefont {B.}~\bibnamefont
  {Carpenter}}, \bibinfo {author} {\bibfnamefont {A.}~\bibnamefont {Gelman}},
  \bibinfo {author} {\bibfnamefont {M.~D.}\ \bibnamefont {Hoffman}}, \bibinfo
  {author} {\bibfnamefont {D.}~\bibnamefont {Lee}}, \bibinfo {author}
  {\bibfnamefont {B.}~\bibnamefont {Goodrich}}, \bibinfo {author}
  {\bibfnamefont {M.}~\bibnamefont {Betancourt}}, \bibinfo {author}
  {\bibfnamefont {M.}~\bibnamefont {Brubaker}}, \bibinfo {author}
  {\bibfnamefont {J.}~\bibnamefont {Guo}}, \bibinfo {author} {\bibfnamefont
  {P.}~\bibnamefont {Li}}, \ and\ \bibinfo {author} {\bibfnamefont
  {A.}~\bibnamefont {Riddell}},\ }\href {\doibase 10.18637/jss.v076.i01}
  {\bibfield  {journal} {\bibinfo  {journal} {Journal of Statistical Software}\
  }\textbf {\bibinfo {volume} {76}},\ \bibinfo {pages} {1} (\bibinfo {year}
  {2017})}\BibitemShut {NoStop}%
\bibitem [{\citenamefont
  {Peixoto}(2014{\natexlab{b}})}]{peixoto2014hierarchical}%
  \BibitemOpen
  \bibfield  {author} {\bibinfo {author} {\bibfnamefont {T.~P.}\ \bibnamefont
  {Peixoto}},\ }\href {\doibase 10.1103/physrevx.4.011047} {\bibfield
  {journal} {\bibinfo  {journal} {Physical Review X}\ }\textbf {\bibinfo
  {volume} {4}} (\bibinfo {year} {2014}{\natexlab{b}}),\
  10.1103/physrevx.4.011047}\BibitemShut {NoStop}%
\bibitem [{\citenamefont {Karrer}\ and\ \citenamefont
  {Newman}(2011)}]{karrer_stochastic_2011}%
  \BibitemOpen
  \bibfield  {author} {\bibinfo {author} {\bibfnamefont {B.}~\bibnamefont
  {Karrer}}\ and\ \bibinfo {author} {\bibfnamefont {M.~E.~J.}\ \bibnamefont
  {Newman}},\ }\href {\doibase 10.1103/PhysRevE.83.016107} {\bibfield
  {journal} {\bibinfo  {journal} {Physical Review E}\ }\textbf {\bibinfo
  {volume} {83}},\ \bibinfo {pages} {016107} (\bibinfo {year}
  {2011})}\BibitemShut {NoStop}%
\bibitem [{\citenamefont {Peixoto}(2017)}]{peixoto2017nonparametric}%
  \BibitemOpen
  \bibfield  {author} {\bibinfo {author} {\bibfnamefont {T.~P.}\ \bibnamefont
  {Peixoto}},\ }\href {\doibase 10.1103/PhysRevE.95.012317} {\bibfield
  {journal} {\bibinfo  {journal} {Phys. Rev. E}\ }\textbf {\bibinfo {volume}
  {95}},\ \bibinfo {pages} {012317} (\bibinfo {year} {2017})}\BibitemShut
  {NoStop}%
\bibitem [{\citenamefont {Peixoto}(2014{\natexlab{c}})}]{peixoto2014efficient}%
  \BibitemOpen
  \bibfield  {author} {\bibinfo {author} {\bibfnamefont {T.~P.}\ \bibnamefont
  {Peixoto}},\ }\href {\doibase 10.1103/PhysRevE.89.012804} {\bibfield
  {journal} {\bibinfo  {journal} {Phys. Rev. E}\ }\textbf {\bibinfo {volume}
  {89}},\ \bibinfo {pages} {012804} (\bibinfo {year}
  {2014}{\natexlab{c}})}\BibitemShut {NoStop}%
\end{thebibliography}%

\newpage
\appendix

\section*{Results on binary network}

\begin{figure}[h!]
    \centering
    \includegraphics[width=\linewidth]{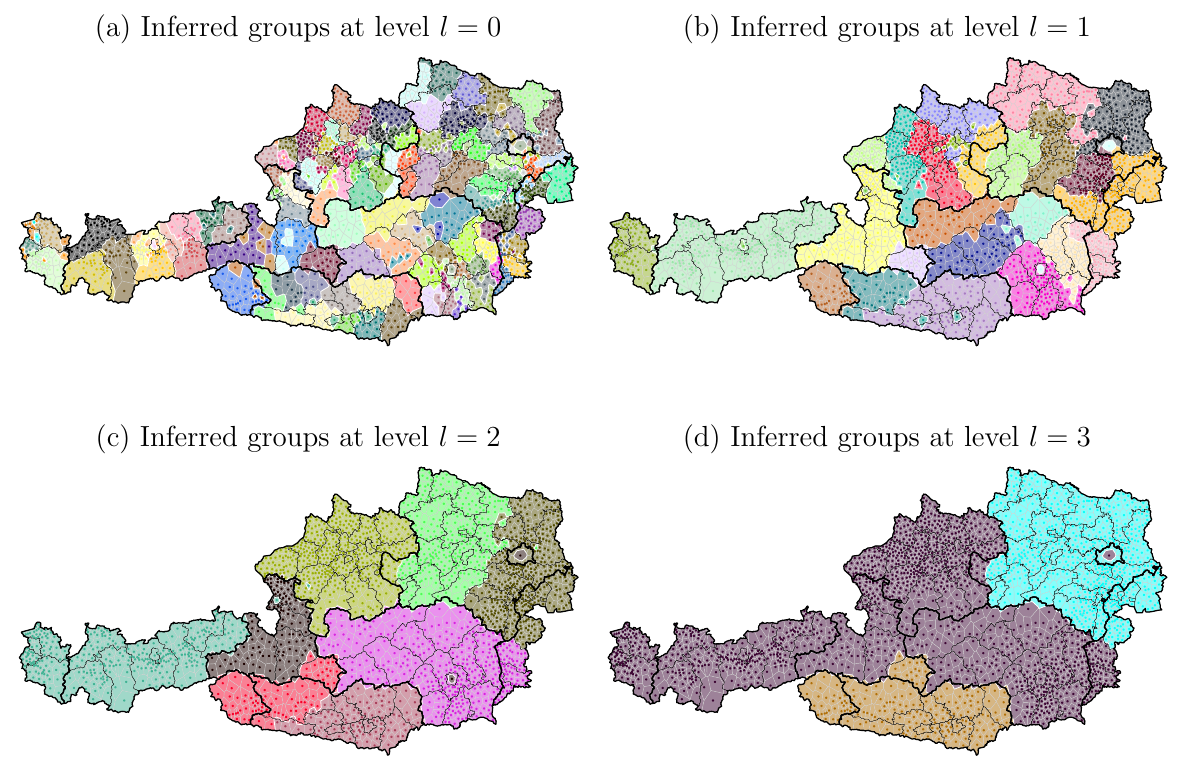}
    \caption{Inferred groups with the SBM for 2013, considering the binarized
      network, for different hierarchical levels $l$. Thick black lines indicate
      federal states boundaries, thin black lines denote districts borders. }
    \label{fig:partition-bin}
\end{figure}

In Fig.~\ref{fig:partition-bin} we show the results equivalent to
Fig.~\ref{fig:partition}, but considering only a binarized version of the
network, characterized by the adjacency matrix
\begin{equation}
A_{ij}= \begin{cases}1, \text{ if  } x_{ij} > 0 \\ 0, \text { otherwise.}\end{cases}
\end{equation}

\end{document}